\documentclass[longauth]{aa}  
\usepackage{ulem}
\usepackage{graphicx}
\usepackage{txfonts}
\usepackage{xcolor}
\usepackage{amsmath, amssymb} 
\usepackage[colorlinks=true,linkcolor=red,anchorcolor=blue,citecolor=blue,filecolor=black,menucolor=black,runcolor=black,urlcolor=blue]{hyperref}

\begin{document}

    \title{Extremely UV-bright starbursts at the end of cosmic reionization}

    \author{R. Marques-Chaves\inst{\ref{inst:Geneva}} \thanks{Rui.MarquesCoelhoChaves@unige.ch} 
    \and D.~Schaerer\inst{\ref{inst:Geneva},\ref{inst:CNRS}}
    \and M.~Dessauges-Zavadsky\inst{\ref{inst:Geneva}}
    \and J.~\'Alvarez-M\'arquez\inst{\ref{inst:CAB}}
    \and T.~Hashimoto\inst{\ref{inst:Tsukuba},\ref{inst:TCHoU}}
    \and L.~Colina\inst{\ref{inst:CAB}}
    \and A.K.~Inoue\inst{\ref{inst:SASE},\ref{inst:Waseda}}
    \and C.~Blanco-Prieto\inst{\ref{inst:CAB}} 
    \and Y.~Nakazato\inst{\ref{inst:DoPTokio},\ref{inst:FLATIRON}}
    \and L.~Costantin\inst{\ref{inst:CAB}}
    \and S.~Arribas \inst{\ref{inst:CAB}}
    \and T.J.L.C.~Bakx\inst{\ref{inst:Chalmers}}
    \and D.~Ceverino\inst{\ref{inst:UAM},\ref{inst:CIAFF}}
    \and A.~Crespo G\'omez\inst{\ref{inst:STScI},\ref{inst:CAB}}
    \and Y.~Fudamoto\inst{\ref{inst:Chiba}}
    \and M.~Hagimoto\inst{\ref{inst:Nagoya}} 
    \and A.~Hamada\inst{\ref{inst:Tsukuba}}
    \and Y.~Matsuoka\inst{\ref{inst:RCSCE}}
    \and K.~Mawatari\inst{\ref{inst:Waseda},\ref{inst:DPAP}}
    \and M.~Onoue\inst{\ref{inst:Waseda2},\ref{inst:Kavli}}
    \and W.~Osone\inst{\ref{inst:Tsukuba}}
    \and Y.W.~Ren\inst{\ref{inst:SASE}}
    \and Y.~Sugahara\inst{\ref{inst:Waseda},\ref{inst:SASE}}
    \and Y.~Terui\inst{\ref{inst:Tsukuba}}
    \and N.~Yoshida\inst{\ref{inst:DoPTokio}, \ref{inst:Kavli}, \ref{inst:RTokio}} 
   }

    \institute{Geneva Observatory, Department of Astronomy, University of Geneva, Chemin Pegasi 51, CH-1290 Versoix, Switzerland \label{inst:Geneva}
    \and CNRS, IRAP, 14 Avenue E. Belin, 31400 Toulouse, France \label{inst:CNRS}
    \and Centro de Astrobiolog\'{\i}a (CAB), CSIC-INTA, Ctra. de Ajalvir km 4, Torrej\'on de Ardoz, E-28850, Madrid, Spain \label{inst:CAB}   
    \and Division of Physics, Faculty of Pure and Applied Sciences, University of Tsukuba, Tsukuba, Ibaraki 305-8571, Japan \label{inst:Tsukuba}
    \and Tomonaga Center for the History of the Universe (TCHoU), Faculty of Pure and Applied Sciences, University of Tsukuba, Tsukuba, Ibaraki 305-8571, Japan \label{inst:TCHoU}
    \and Department of Physics, School of Advanced Science and Engineering, Faculty of Science and Engineering, Waseda University, 3-4-1 Okubo, Shinjuku, Tokyo 169-8555, Japan \label{inst:SASE}
    \and Waseda Research Institute for Science and Engineering, Faculty of Science and Engineering, Waseda University, 3-4-1 Okubo, Shinjuku, Tokyo 169-8555, Japan \label{inst:Waseda}
    \and Department of Physics, The University of Tokyo, 7-3-1 Hongo, Bunkyo, Tokyo 113-0033, Japan \label{inst:DoPTokio}
    \and Center for Computational Astrophysics, Flatiron Institute, 162 5th Avenue, New York, NY 10010 \label{inst:FLATIRON}
    \and Department of Space, Earth and Environment, Chalmers University of Technology, SE-412 96 Gothenburg, Sweden \label{inst:Chalmers}  
    \and Space Telescope Science Institute (STScI), 3700 San martin Drive, Baltimore, MD 21218, USA \label{inst:STScI}
    \and Departamento de Fisica Teorica, Modulo 8, Facultad de Ciencias, Universidad Autonoma de Madrid, 28049 Madrid, Spain \label{inst:UAM}
    \and  CIAFF, Facultad de Ciencias, Universidad Autonoma de Madrid, 28049 Madrid, Spain \label{inst:CIAFF}
    \and Center for Frontier Science, Chiba University, 1-33 Yayoi-cho, Inage-ku, Chiba 263-8522, Japan \label{inst:Chiba}
    \and Department of Physics, Graduate School of Science, Nagoya University, Nagoya 464-8602, Japan \label{inst:Nagoya} 
    \and Research Center for Space and Cosmic Evolution, Ehime University, Matsuyama, Ehime 790-8577, Japan  \label{inst:RCSCE}
    \and Department of Pure and Applied Physics, School of Advanced Science and Engineering, Faculty of Science and Engineering, Waseda University, 3-4-1 Okubo, Shinjuku, Tokyo 169-8555, Japan \label{inst:DPAP}
    \and Waseda Institute for Advanced Study (WIAS), Waseda University, 1-21-1, Nishi-Waseda, Shinjuku, Tokyo 169-0051, Japan; Center for Data Science, Waseda University, 1-6-1, Nishi-Waseda, Shinjuku, Tokyo 169-0051, Japan \label{inst:Waseda2}
    \and  Kavli Institute for the Physics and Mathematics of the Universe (WPI), UT Institute for Advanced Study, The University of Tokyo, Kashiwa, Chiba 277-8583, Japan \label{inst:Kavli}
    \and Research Center for the Early Universe, School of Science, The University of Tokyo, 7-3-1 Hongo, Bunkyo, Tokyo 113-0033, Japan \label{inst:RTokio}  }

   \date{Received ; accepted}


 

\abstract{We present a study of 27 very UV-bright ($-22.0 \lesssim M_{\rm UV} \lesssim -24.4$) star-forming galaxies at $z \sim 6$, identified in the Subaru High-$z$ Exploration of Low-Luminosity Quasars (SHELLQs) survey. Stacking their rest-frame UV spectra reveals a prominent N\,{\sc v} $\lambda$1240 P-Cygni feature, consistent with very young ($\sim$6 Myr) stellar populations dominated by massive and hot stars. UV-bright galaxies in the reionization epoch are thus powerful and efficient ionizing sources, with an average ionizing photon production efficiency of $\log(\xi_{\rm ion}/{\rm Hz\,erg^{-1}}) = 25.54^{+0.09}_{-0.12}$. 
For one representative source, J0217–0208 at $z=6.204$ ($M_{\rm UV} = -23.4$), we analyzed available JWST/NIRCam and NIRSpec observations. 
Its spectral energy distribution indicates a young ($\sim 5$ Myr) starburst with a stellar mass of $10^{9}\,M_{\odot}$ and a high specific star formation rate ($\sim 100$ Gyr$^{-1}$). Together with its very compact NIRCam-measured size ($r_{\rm eff} \simeq 260$\,pc), this corresponds to stellar mass and star formation rate surface densities $\approx 100\times$ higher than those of typical galaxies at comparable redshifts. 
NIRSpec spectroscopy further reveals strong nebular emission, for which we derive a high electron density ($n_{\rm e} \simeq 10^{3}$\,cm$^{-3}$), a metallicity 12+$\log({\rm O/H}) = 8.20 \pm 0.11$ (from the direct method) and a super-solar N/O ratio (log(N/O)$\simeq -0.30$). Furthermore, J0217–0208 shows broad components in several rest-optical emission lines, indicating powerful ionized outflows. From the Balmer decrement, these outflows appear heavily obscured ($E(B-V)_{\rm out} \simeq 0.6$), in contrast to the nearly dust-free stellar continuum ($E(B-V)_{\star} = 0.01 \pm 0.01$) obtained from its steep UV slope, $\beta_{\rm UV} \simeq -2.6$. Combined with ALMA detections of a massive ($M_{\rm dust} \simeq 2 \times 10^{8}\,M_{\odot}$), extended ($\sim$1.4 kpc), and cold ($T_{\rm dust} \simeq 25$ K) dust reservoir, these findings point to dusty, feedback-driven outflows carrying and pushing dust well beyond the stellar core and likely boosting the observed UV luminosity.
Taken together, our results suggest that UV-bright galaxies at high redshift represent short-lived but extreme phases of rapid stellar mass growth, efficient ionizing photon production, and strong feedback. The extreme properties of J0217–0208, such as supersolar N/O, steep UV slope, compact size, and very high surface densities, closely mirror those of the brightest galaxies at $z>10$, suggesting a shared evolutionary pathway.}

\keywords{Galaxies: starburst -- Galaxies: high-redshift -- Cosmology: dark ages, reionization, first stars}
\titlerunning{EoR UV-bright starbursts}
\maketitle

\section{Introduction}

UV-bright star-forming galaxies, sources located at the bright end of the UV luminosity function, were once considered extremely rare at any redshift, including during the Epoch of Reionization (EoR, $6 < z < 16$). However, the advent of the \textit{James Webb} Space Telescope (JWST) has dramatically changed this view by revealing an unexpectedly large number of UV-luminous sources at $z \gtrsim 10$ \citep[e.g.,][]{ArrabalHaro2023, Carniani2024}. Some of these systems exhibit properties rarely observed at lower redshift, including unusual abundance patterns inferred from intense rest-UV N\,{\sc iv}] $\lambda$1486 and N\,{\sc iii}] $\lambda$1750 emission lines \citep[][]{Bunker2023, Castellano2024, Naidu2025}, as well as compact morphologies and high stellar mass and star-formation surface densities \citep{Schaerer2024}. Regardless of their intrinsic properties, their observed number densities exceed predictions from pre-JWST models of galaxy formation by nearly an order of magnitude \citep{Kannan2023, Lovell2023MNRAS}, highlighting a striking tension with theoretical expectations.
\looseness=-1

Hints of this discrepancy were already present before JWST. Wide-field imaging surveys had uncovered an apparent excess of bright galaxies at $z \gtrsim 8$ relative to Schechter-function extrapolations, as well as a surprisingly slow evolution of the bright end of the UV luminosity function at these redshifts \citep[e.g.,][]{oesch2016, Morishita2018, stefanon2019, Bowler2020, harikane2022b}. JWST has now confirmed and extended these trends to even earlier epochs, firmly establishing the existence of an overabundant population of UV-bright systems in the first few hundred Myr of cosmic time.
\looseness=-1

Several explanations have been put forward to account for this tension. One possibility is that early galaxies formed stars with much higher efficiencies than typically assumed and measured locally, aided by the dense, metal-poor conditions at early epochs \citep[][see also: \citealt{Boylan-Kolchin2025, Renzini2025}]{Dekel2023, Ceverino2024, Li2024}. Alternatively, a non-standard stellar initial mass function (IMF) biased toward (very) massive stars could boost the UV luminosity-to-mass ratio \citep[e.g.,][]{Trinca2024, Hutter2025}. Other scenarios include strong radiation-driven outflows that rapidly expel dust, thereby reducing attenuation and enhancing the UV output \citep{Ferrara2023, Ziparo2023}, or stochastic star-formation histories and very young stellar ages that can transiently elevate the UV luminosity \citep[e.g.,][]{Mason2023, Ciesla2024, Kravtsov2024, Donnan2025}. While these frameworks differ fundamentally, they all invoke extreme conditions rarely observed at lower redshifts.
\looseness=-1

\begin{figure*}
  \centering
  \includegraphics[width=0.98\textwidth]{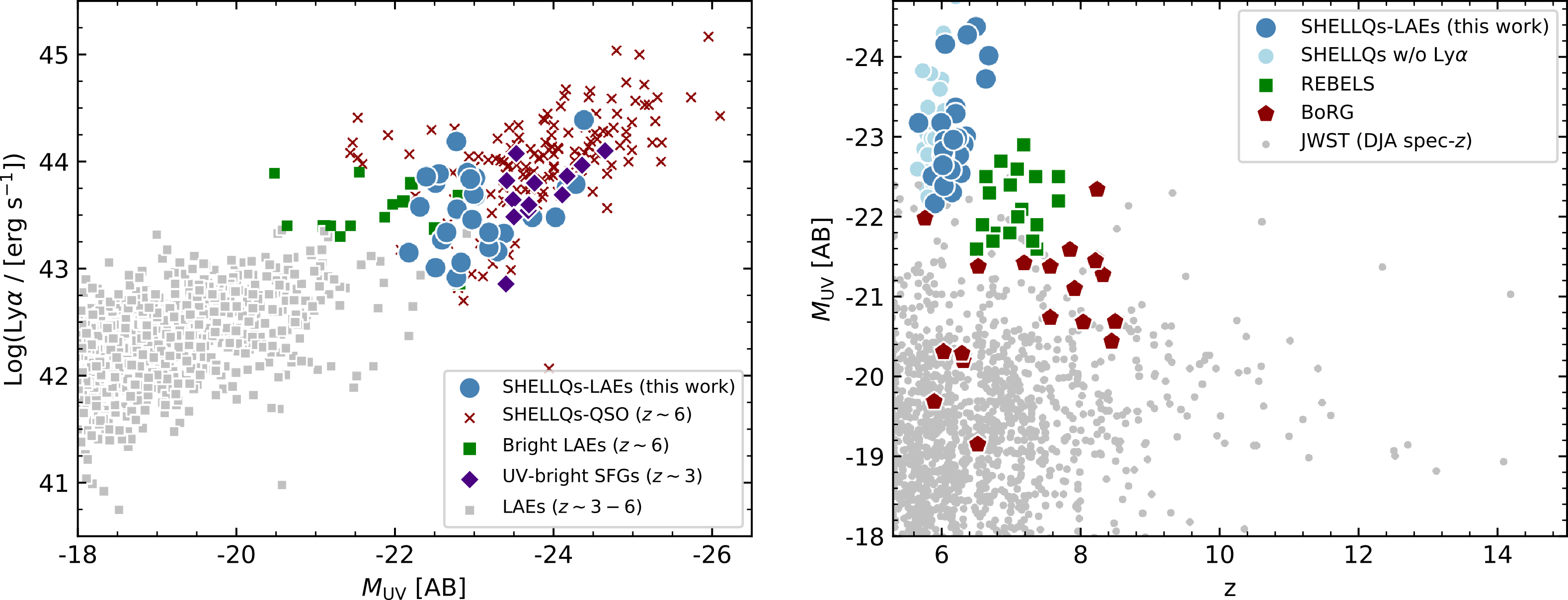}
  \caption{Distribution of Ly$\alpha$ luminosity ($L_{\rm Ly\alpha}$), UV absolute magnitude ($M_{\rm UV}$), and redshift ($z$) for the SHELLQs-LAEs analyzed in this work (blue circles). Left panel: $L_{\rm Ly\alpha}$–$M_{\rm UV}$ plane comparing SHELLQs LAEs with other galaxy populations with spectroscopic redshifts: SHELLQs-QSOs at $z\sim6$ (red crosses; \citealt{matsuoka2016, matsuoka2018a, matsuoka2018b, matsuoka2019, matsuoka2022, Matsuoka2025b}), bright LAEs at $z\sim6$ (green squares; \citealt{ouchi2009, sobral2015, matthee2017, Shibuya2018PASJ...70S..15S, Marconcini2024}), UV-bright star-forming galaxies at $z\sim3$ from SDSS/BOSS (violet diamonds; \citealt{marques2020, marques2020b, marques2021, marques2022, Dessauges-Zavadsky2025}), and more typical LAEs at $z\sim3$–6 (grey squares; \citealt{Kerutt2022A&A...659A.183K}). Right panel: $M_{\rm UV}$–$z$ distribution of the SHELLQs LAEs together with UV-bright galaxies from SHELLQs lacking or showing weak Ly$\alpha$ emission (light blue circles; \citealt{matsuoka2016, matsuoka2018a, matsuoka2018b, matsuoka2019, matsuoka2022}), and sources from the REBELS (green; \citealt{bouwens2022}) and BoRG (red; \citealt{roberts2024_borg}) surveys, as well as a compilation of spectroscopically confirmed $z_{\rm spec}>5$ objects from the DAWN/DJA archive (grey).}
  \label{fig1}
\end{figure*}

\looseness=-1

Crucially, and except for a few sources \citep[e.g.,][]{Bunker2023, Carniani2024, Castellano2024, Fudamoto2024_GNz11_NOEMA, AlvarezMarquez2025_GNz11, Schouws2025, Zavala2025_GHz12_MIRI}, most $z > 10$ UV-bright galaxies remain too faint for detailed, high signal-to-noise multi-wavelength studies. As a result, their stellar populations, interstellar medium conditions, and dust properties remain poorly constrained, leaving the physical drivers of their remarkable luminosities an open question. In this sense, lower-$z$ analogs may serve as ideal laboratories to investigate and potentially establish the conditions and properties of these extreme systems. In particular, identifying the UV-brightest systems at the highest accessible redshifts is crucial, as the aforementioned mechanisms may be most pronounced there and the physical conditions may remain closest to those of the early Universe.
\looseness=-1

In this work, we present a study of a large sample of UV-bright galaxies ($-22.0 \lesssim M_{\rm UV} \lesssim -24.4$) at $z \sim 6$, identified in the Subaru High-$z$ Exploration of Low-Luminosity Quasars survey (SHELLQs; \citealt{matsuoka2016, matsuoka2018a, matsuoka2018b, matsuoka2019, matsuoka2022, Matsuoka2025b}). This work is organized as follows. In Section \ref{section2}, we investigate the average rest-UV properties of these sources using ground-based spectroscopic observations. Section \ref{section3} presents JWST observations and corresponding analysis of one UV-bright galaxy at $z=6.2$.  The discussion of the results is presented in Section \ref{section4}, and, finally, we present the summary of our main findings in Section \ref{section5}. Throughout this work, we use a concordance cosmology with $\Omega_{\rm m} = 0.274$, $\Omega_{\Lambda} = 0.726$, and $H_{0} = 70$\,km\,s$^{-1}$\,Mpc$^{-1}$. Magnitudes are given in the AB system.
\looseness=-1


\section{Average properties of UV-bright EoR galaxies}\label{section2}

\subsection{Sample and observations}\label{selection}

SHELLQs is a wide-field survey designed to identify low-luminosity QSOs at $6 \lesssim z \lesssim 7$ using ground-based imaging data from the Hyper Suprime-Cam (HSC) Subaru Strategic Program. As detailed in \cite{matsuoka2022}, candidates are selected based on the dropout technique in the $i$- or $z$-bands and their point-like morphologies in HSC images. Their redshifts are confirmed through optical spectroscopic follow-up observations with the Subaru and the Gran Telescopio Canarias (GTC) telescopes, using Ly$\alpha$ emission or the Ly$\alpha$ continuum break.
\looseness=-1

Among more than 160 high-$z$ sources identified so far within SHELLQs, some exhibit strong ($L_{\rm Ly\alpha} > 10^{43}$\,erg\,s$^{-1}$) and narrow (FWHM < 1000 km s$^{-1}$) Ly$\alpha$ emission. Since similarly high Ly$\alpha$ luminosities are also seen at lower redshifts (e.g., at cosmic noon) and sometimes associated with AGN activity (e.g., \citealt{konno2016}), SHELLQs conservatively classifies objects with $L_{\rm Ly\alpha} > 10^{43}$\,erg\,s$^{-1}$ as AGNs, either obscured AGNs or mini/weak broad absorption line (BAL) QSOs \citep{matsuoka2018a, matsuoka2019, matsuoka2022}. Nonetheless, as shown in the left panel of Figure~\ref{fig1}, several star-forming galaxies at intermediate and high redshifts also exhibit Ly$\alpha$ luminosities well exceeding $L_{\rm Ly\alpha} > 10^{43}$\,erg\,s$^{-1}$ without any evidence of AGN \citep[e.g.,][to name a few]{ouchi2009, sobral2015, matthee2017, Shibuya2018PASJ...70S..15S, marques2020, marques2020b, marques2021, marques2022, Marconcini2024, Torralba-Torregrosa2024, Dessauges-Zavadsky2025}.
\looseness=-1

In this study, we reanalyze the spectra and classification of UV-bright SHELLQs sources with strong and narrow Ly$\alpha$ emission (hereafter SHELLQs-LAEs). To focus on the most UV-bright sources, we selected objects with $M_{\rm UV} \leq -22$ measured from ground-based images), $L_{\rm Ly\alpha} > 10^{43}$\,erg\,s$^{-1}$, and FWHM < 1000 km s$^{-1}$. This subsample consists of 27 sources, with a mean and scatter absolute UV magnitude of $M_{\rm UV} = -23.03 \pm 0.58$, $z=6.15\pm0.22$, and log($L_{\rm Ly\alpha}/ \rm erg$\,s$^{-1}) = 43.57 \pm 0.36$. 

Fully reduced optical spectra obtained with Subaru/FOCAS and GTC/OSIRIS are retrieved from the SHELLQs webpage\footnote{\url{https://cosmos.phys.sci.ehime-u.ac.jp/~yk.matsuoka/shellqs.html}}. A detailed description of the observations and data reduction is provided in the main SHELLQs survey papers \citep{matsuoka2016, matsuoka2018a, matsuoka2018b, matsuoka2019, matsuoka2022}. Briefly, the observations were conducted between November 2015 and August 2021, with typical total on-source exposure times of 30-120\,min. The reduced one-dimensional spectra reach $1\sigma$ depths of $\simeq(4$–$10)\times10^{-19}$\,erg\,s$^{-1}$\,cm$^{-2}$\,\AA$^{-1}$ over the rest-frame wavelength range $\lambda_{\rm rest}=1250$–1400\,\AA. Within this region, the average signal-to-noise ratio is $\sim0.8$ per pixel ($\simeq 1.5$\,\AA\,pix$^{-1}$), corresponding to an integrated S/N of $\sim20$. All the rest-UV spectra used in this work are presented in the original SHELLQs survey papers.

Appendix \ref{appendix1} and Table \ref{table_sample} list their coordinates and general properties obtained by the SHELLQs collaboration. The UV absolute magnitudes of SHELLQs-LAEs are represented as dark blue circles in Figure \ref{fig1}. For comparison, the right panel of Figure \ref{fig1} also shows other UV-bright LBGs selected from SHELLQs (but without or weak Ly$\alpha$ emission; in light blue; \citealt{matsuoka2019}) and other surveys targeting UV-bright star-forming galaxies at $z\gtrsim 6$ such as REBELS \citep[green;][]{bouwens2022, Rowland2025} and BoRG \citep[red;][]{roberts2024_borg}, along with a compilation of sources with spectroscopic redshifts above $z_{\rm spec}>5$ obtained with NIRSpec/JWST \citep[grey, from DAWN/DJA archive;][]{Heintz2024_DJA, deGraaff2025_DJA}.
\looseness=-1

\begin{figure*}
  \centering
  \includegraphics[width=0.72\textwidth]{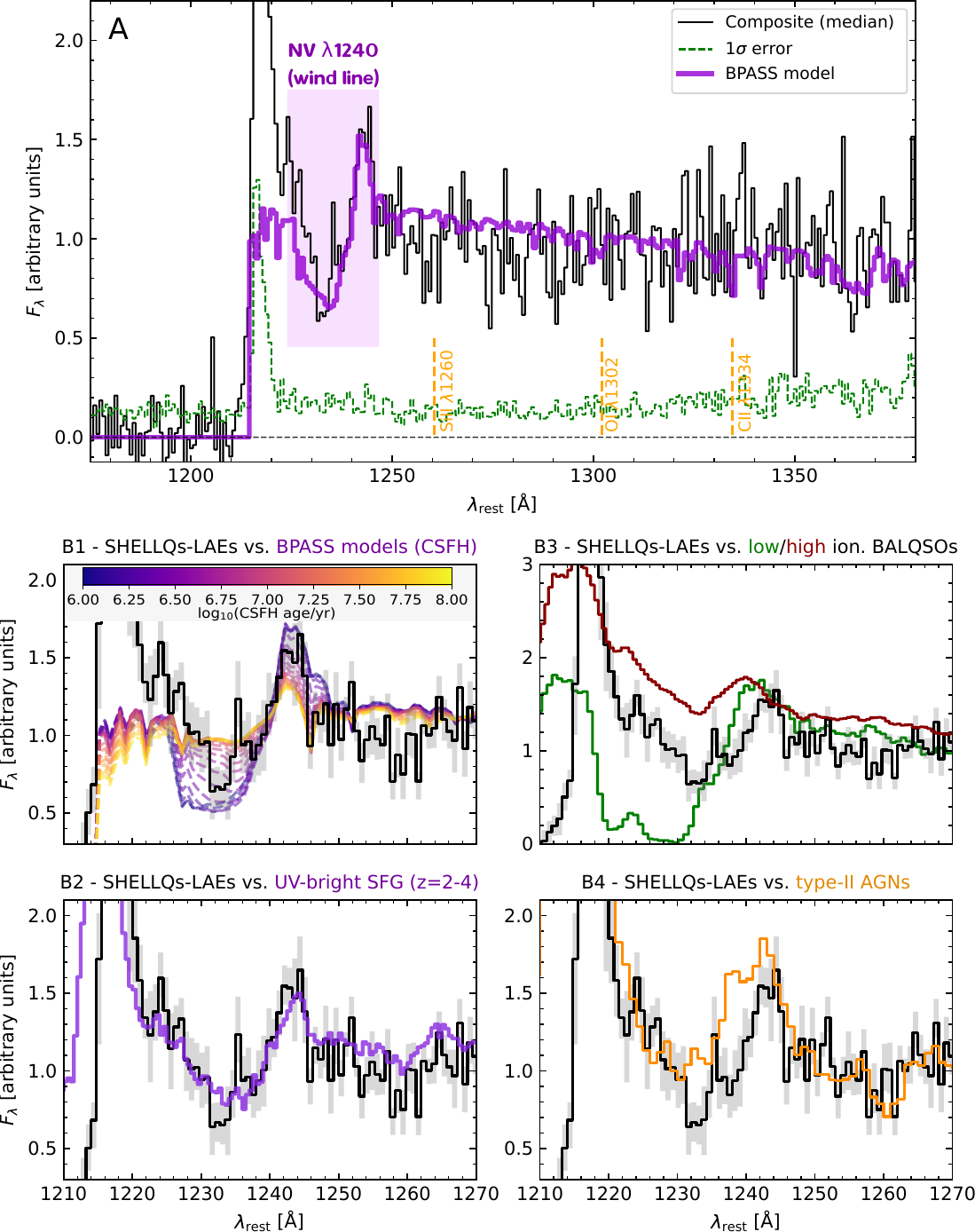}
      \caption{Stacked rest-UV spectrum of UV-bright SHELLQs-LAEs. Panel A shows the median (black) stacked spectrum and $1\sigma$ error (green) of SHELLQs-LAEs, which presents strong Ly$\alpha$ emission and a well-developed N~{\sc v} $\lambda$1240 P-Cygni profile (violet-shaded area). The location of low-ionization ISM absorption lines (Si~{\sc ii}, O~{\sc i}, and C~{\sc ii}) are also marked with orange dashed lines. The BPASS stellar population synthesis model that best reproduces the observed N~{\sc v} profile is shown in violet and is characterized by a continuous star formation with an age of $6.3^{+9}_{-3}$\,Myr, assuming $Z/Z_{\odot}=0.3$ with the \cite{chabrier2003} IMF and $E(B-V)=0$. Bottom panels (B1-B4) show a zoom-in to the N~{\sc v} P-Cygni observed in SHELLQs-LAEs (black) and a comparison with B1: BPASS synthetic models with different ages ($1-100$\,Myr, color-coded); B2: UV-bright galaxies at $z=2-4$ \citep[violet;][]{Upadhyaya2024}; B3: low- and high-ionization BALQSOs \citep[green and red, respectively;][]{Brotherton2001}; and B4: type II AGNs \citep[orange;][]{hainline2011b}.  }
  \label{fig2}
\end{figure*}

\subsection{Composite spectrum}
\looseness=-1

To investigate the representative spectral properties of the SHELLQs-LAEs, we construct a composite rest-frame UV spectrum by stacking the individual spectra. While Ly$\alpha$ redshifts are available for all sources, accurate systemic redshifts are crucial for aligning spectra before stacking. However, these are only known for a few SHELLQs-LAEs from recent JWST spectroscopy \citep{Matsuoka2025}: J0217-0208 ($z=6.204$, further discussed in Section \ref{section3}), J0853+0139 ($z=6.007$), J0905+0300 ($z=6.273$), and J1416+0015 ($z=6.027$). For the remaining sources, we estimate the systemic redshifts using the empirical corrections from \cite{verhamme2018} using the Ly$\alpha$ emission peak and line width derived from optical spectroscopy, which account for radiative transfer effects.

Each spectrum is then de-redshifted using the estimated systemic redshifts and resampled onto a common spectral grid with a resolution of $\Delta \lambda = 1.5$\,\AA. We also rescaled the flux of every spectrum to a common flux density to match the mean absolute UV magnitude of the sample ($M_{\rm UV} \simeq -23.0$).
The composite spectrum is obtained by taking the median of the resampled spectra at each wavelength bin, minimizing the contamination in the final stack from strong sky-line residuals. 

For the stacked spectrum error, we first compute the formal per-pixel uncertainty from inverse-variance weighting of the contributing spectra. However, this estimate does not account for intrinsic sample variance or correlated noise introduced by resampling onto a common wavelength grid. We therefore estimate the final uncertainties using bootstrap resampling.
We generate 1000 realizations of the composite spectrum by randomly drawing spectra with replacement and recomputing the median stack for each realization. The $1\sigma$ uncertainty at each wavelength is taken as the standard deviation of the bootstrap distribution. This approach incorporates both measurement noise and intrinsic spectral diversity within the population.
\looseness=-1

Figure \ref{fig2}-A presents the median stacked spectrum. It exhibits strong ($EW_{0} = 19\pm3$\,\AA) and narrow ($\rm FWHM = 612 \pm 56$\,km\,s$^{-1}$) Ly$\alpha$ emission and a strong P-Cygni profile around N~{\sc v}~$\lambda$1240. Other spectral features are tentatively detected, including the low-ionization absorption line in Si~{\sc ii}~$\lambda$1260. To assess the robustness of the stack, we also computed the mean composite spectra , finding consistent results, but with lower SNR.
\looseness=-1

\subsection{Nature of UV-bright galaxies: very young stellar populations}\label{nature-shellqs}
\looseness=-1

The most striking spectral feature observed in the median stacked spectrum shown in Figure \ref{fig2} is the P-Cygni profile around N~{\sc v}~$\lambda$1240.
This feature, characterized by blueshifted absorption and redshifted emission, is typical of the most massive O-type stars and originates from their strong stellar winds (c.f., \citealt{roman2020}). The strength of the N{\sc v} P-Cygni profile is thus sensitive to the age of the stellar population but, as highlighted in \cite{chisholm2019}, it is only weakly dependent on stellar metallicity, in contrast to other wind lines such as Si~{\sc iv} $\lambda$1400 and C~{\sc iv}~$\lambda$1550 whose strength is more dependent on the metallicity. As such, strong N~{\sc v} P-Cygni is observed (and expected) in systems dominated by very young stellar populations \citep[e.g.,][]{leitherer2018, chisholm2019, Berg2022, Smith2023, Upadhyaya2024, Marques-Chaves2026arXiv260421493M}. For older or more evolved stellar populations, the N~{\sc v} P-Cygni becomes significantly weaker or is no longer discernible, as seen in the spectra of more evolved galaxies \citep[e.g.,][]{shapley2003, Sugahara2019}.
\looseness=-1

To investigate the origin of the N~{\sc v} P-Cygni feature in our stacked spectrum, we fit its observed profile (from $\lambda_{\rm rest} = 1225-1248$\,\AA) using predictions from stellar population synthesis models. We use the Binary Population and Spectral Synthesis (BPASS v2.1; \citealt{eldridge2017}) code, assuming a constant star formation history (CSFH) and a \cite{chabrier2003} IMF. Because the composite spectrum combines 27 galaxies that are unlikely to share identical star formation histories, a CSFH provides a neutral time-averaged description without imposing a specific functional form. We adopt a stellar metallicity of $Z_{\star}=0.006$ (or $Z=0.3Z_{\odot}$, assuming $Z_{\odot}=0.02$, see also Section~\ref{density}). The observed N~{\sc v} profile is well reproduced by stellar populations with an age of $\simeq 6$\,Myr (Figure~\ref{fig2}-A). Figure~\ref{fig2}-B1 shows a zoom-in of the observed N~{\sc v} feature in our stacked spectrum, compared with BPASS models spanning a wide range of ages ($1-100$\,Myr, color-coded). Despite the relatively low SNR, our data suggest a relatively well-constrained age of $6.3^{+9}_{-3}$\,Myr. Ages shorter than $\simeq 3$\,Myr or longer than $\simeq 15$\,Myr are significantly disfavored, as they either overpredict or underpredict the observed strength of the N~{\sc v} P-Cygni feature.
\looseness=-1

It is noteworthy that similar results have been found for UV-bright galaxies at intermediate redshifts ($z\sim 2-4$ and $M_{\rm UV} \leq -23$), which are so far the only star-forming galaxies known to be as bright as the SHELLQs-LAEs. Deep rest-frame UV spectroscopy of these $z \sim 2-4$ galaxies has revealed intense stellar wind lines, including prominent N~{\sc v} P-Cygni, from which ages of $\lesssim 10$\,Myr and high specific star formation rates ($\rm sSFR = SFR/M_{\star} \sim 100$\,Gyr$^{-1}$) were inferred \citep[][]{marques2020b, marques2021, marques2022, Upadhyaya2024, Dessauges-Zavadsky2025, Marques-Chaves2024b}. Figure~\ref{fig2}-B2 compares the N~{\sc v} profile of the SHELLQs-LAEs with that of the stacked spectrum of similarly UV-bright galaxies at $z \sim 2-4$ presented in \citet{Upadhyaya2024}, showing a great similarity between the two.
\looseness=-1

The N~{\sc v} P-Cygni feature in SHELLQs-LAEs was also noted by \cite{matsuoka2019}, who performed a similar spectral stacking analysis using a slightly smaller sample (see their Fig.\,9). They attributed the observed N~{\sc v} P-Cygni profile to a weak broad absorption line quasar (BALQSO), a type of quasar that shows blueshifted absorption features in its spectrum due to outflowing gas from the central supermassive black hole. 
While the weak BALQSO scenario cannot be ruled out as an explanation for the observed N~{\sc v} profile in the SHELLQs-LAEs, it is unlikely for several reasons. First, the P-Cygni-like features observed in BALQSOs arise from the combination of broad emission from the accretion disk and absorption by outflowing gas along the line of sight, typically spanning a wide range of velocities from a few hundred km\,s$^{-1}$ to relativistic speeds (cf. \citealt{Hamann1999}, for a review).
Given this diversity, it is improbable (but not impossible) that stacking the rest-UV spectra of multiple BALQSOs would yield a N~{\sc v} P-Cygni profile that precisely matches that expected from stellar winds of massive stars. Figure~\ref{fig2}-B3 shows a comparison between the observed N~{\sc v} profile of the SHELLQs-LAEs and those obtained from stacks of low- (green) and high-ionization (red) BALQSOs at $z\sim 1$ \citep{Brotherton2001}. While these BALQSO composites do exhibit P-Cygni-like N~{\sc v} profiles, they differ significantly from the one observed in the SHELLQs-LAEs and from BPASS model predictions. Furthermore, BALQSOs typically show broad Ly$\alpha$ profiles (with FWHM $>1000$\,km\,s$^{-1}$), in contrast to the narrower Ly$\alpha$ line widths observed in SHELLQs-LAEs ($<1000$\,km\,s$^{-1}$).
\looseness=-1

Finally, we investigate the contribution of an obscured AGN (e.g., type II). This is more challenging due to the limited spectral coverage of our stacked spectrum ($\lambda_{\rm rest} \lesssim 1400$\,\AA). However, as shown in Figure~\ref{fig2}-B4 (orange), type-II AGNs typically exhibit narrow and symmetric N~{\sc v} emission, from N~{\sc v} 1238\,\AA{} and 1242\,\AA{ }\citep{hainline2011b, alexandroff2013, Mignoli2019}, which contrasts with blueshifted absorption and redshifted emission observed in SHELLQs-LAEs. Furthermore, a type-II AGN or, more generally, an obscured AGN, would not account for the high UV luminosities observed in the SHELLQs-LAEs.
\looseness=-1

\begin{figure*}
  \centering
  \includegraphics[width=0.98\textwidth]{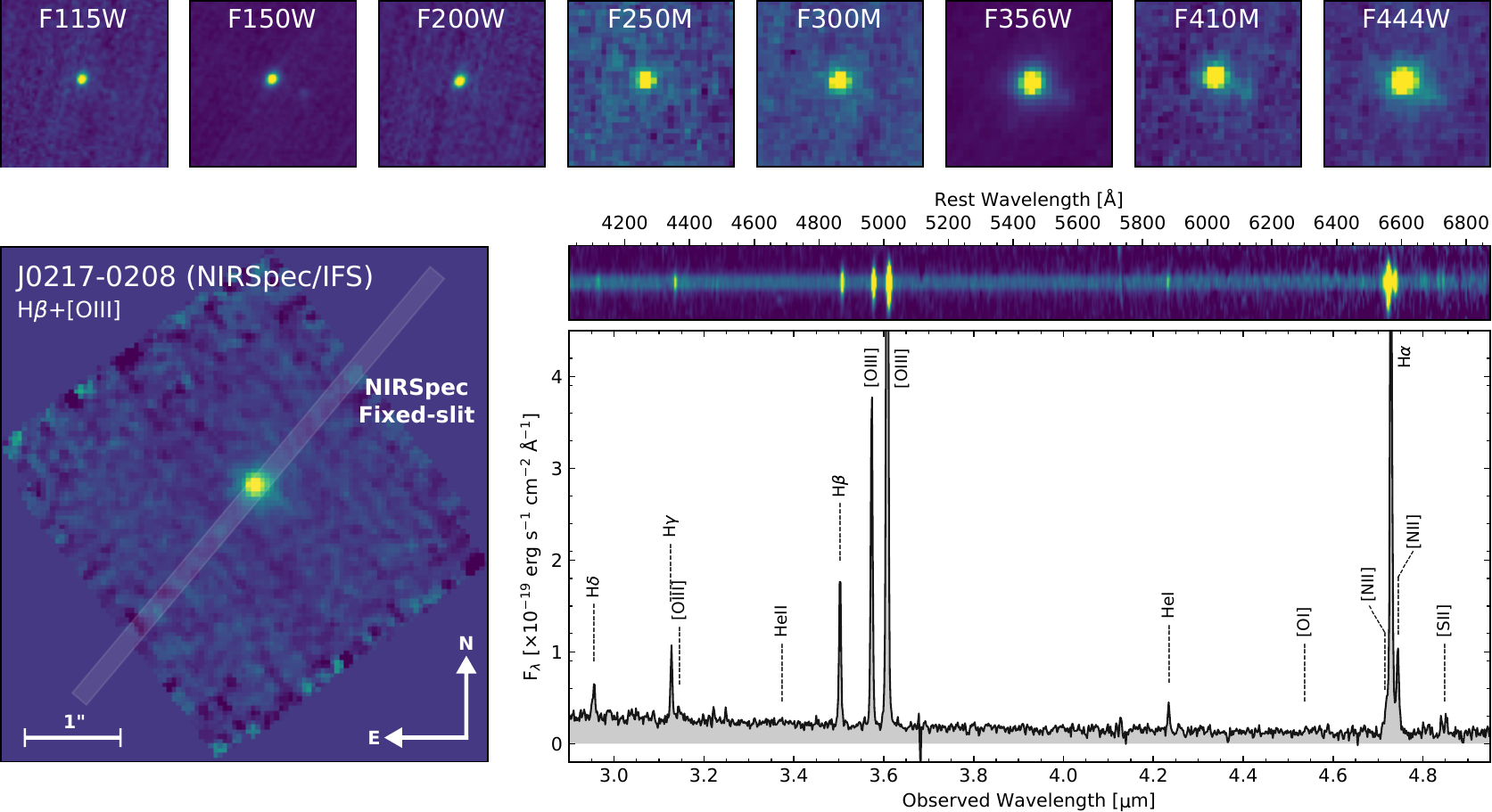}
    \caption{Summary of the JWST observations of J0217-0208 at $z=6.20$. The top panels show JWST/NIRCam images centered on J0217-0208 at increasing wavelengths (from F115W to F444W, left to right, each with a $2^{\prime\prime}\times2^{\prime\prime}$ size). The bottom panels present JWST/NIRSpec spectroscopy: the left panel shows the integrated H$\beta$+[O~{\sc iii}]~$\lambda\lambda$4960,5008 emission from J0217-0208 obtained with NIRSpec/IFS, while the right panel shows the 2D (top) and 1D (bottom) spectra of J0217-0208 obtained with NIRSpec/Fixed-slit. The locations of relevant emission lines are marked with dashed lines.}
  \label{fig3}
\end{figure*}
\looseness=-1

In summary, our results suggest that the UV emission of the UV-bright SHELLQs-LAEs studied here is likely dominated by very young stellar populations, with an average age of $\simeq 6$\,Myr under the assumption of a constant star formation rate. As noted earlier, this does not rule out the presence of AGN-dominated systems within the sample \citep[see e.g.,][]{Onoue2021}, but their overall contribution to the UV or frequency is likely to be small. Deeper follow-up spectroscopy in the rest-UV and/or optical is required to determine the nature of these systems on an individual basis. Such observations are available for five UV-bright SHELLQs-LAEs \citep[J0217-0208, J0853+0139, J0905+0300, J1416+0015, and J2232+0012;][]{Matsuoka2025}, four of which show no evidence of AGN activity, as indicated by the absence of broad ($\rm FWHM > 1000$\,km\,s$^{-1}$) components in the Balmer lines, the non-detection of high-ionization emission lines, and slightly resolved morphologies \citep{Matsuoka2025}. 
\looseness=-1

In the following, we present an analysis of JWST observations of one of these sources, J0217-0208 at $z=6.204$. This galaxy is representative of the SHELLQs-LAE population in terms of UV and Ly$\alpha$ luminosities and is the only object in the sample with multi-wavelength coverage including ALMA, which, as we will show later, provides crucial constraints for a comprehensive understanding of the system.
\looseness=-1

\section{JWST observations of J0217-0208}\label{section3}
\looseness=-1

J0217-0208 (RA: 02:17:21.59; Dec: $-$02:08:52.6) at $z=6.20$ and with $M_{\rm UV}=-23.3$ was first discovered in \cite{matsuoka2018a}, showing a relatively strong Ly$\alpha$ emission (log[$L_{\rm Ly\alpha}/ \rm erg$\,s$^{-1}] = 43.33$) and unresolved profile ($\rm FWHM < 230$\,km\,s$^{-1}$). 
Far-IR ALMA observations of this source were presented in \cite{Harikane2020b}, which reported significant detections in the [O~{\sc iii}]\,88$\mu$m and [C~{\sc ii}]\,158$\mu$m. Dust continuum emission was also significantly detected in J0217-0208 at 160\,$\mu$m and 120\,$\mu$m, for which \cite{Harikane2020b} derived a total IR luminosity $L_{\rm IR} = 1.4\times 10^{11}$\,$L_{\odot}$, and a dust temperature and mass $T_{\rm dust} \simeq25$\,K and $M_{\rm dust} \simeq 2\times10^{8}$\,$M_{\odot}$. Figure \ref{fig3} summarizes the JWST observations of J0217-0208. The properties of J0217-0208 derived in this section are summarized in Table \ref{table1}.
\looseness=-1

\subsection{Observations and data reduction}\label{jwst_observations}

J0217-0208 was observed with JWST in three different GO1 programs: PIDs 1840 \citep[PIs: Alvarez-Marquez and T. Hashimoto;][]{Hashimoto2023, Sugahara2024, Mawatari2025, Usui2025}, 1967 \citep[PI: M. Onoue;][]{Ding2025, Matsuoka2025}, and 1657 \citep[PI: Y. Harikane;][]{harikane2025}. These observations consist of NIRCam images with eight filters and NIRSpec spectroscopy with Fixed-Slit (FS) and Integral Field Spectroscopy (IFS) modes that we now describe. 

J0217-0208 was observed with NIRCam with F115W, F150W, F200W, F250M, F300M, F356W, F410M, and F444W filters, consisting of relatively short exposure times (343 to 601\,s on source), except for F150W and F356W (3264\,s on source). 
NIRCam data were calibrated using a custom strategy based on the JWST calibration pipeline (version 1.17.1; \citealt{Bushouse2025zndo}) with Calibration Reference Data System (CRDS) context 1321. This procedure includes the removal of snowballs and wisps following the method described in \cite{Bagley2023}, as well as a superbackground homogenization with 1/$f$ correction, adopting the approach of \cite{PG2023} and incorporating the improvements presented in \cite{PG2025} and \cite{Ostlin2025}. J0217-0208 is well detected in all NIRcam images, as shown in the top panel of Figure \ref{fig3}. 

NIRSpec observations of J0217-0208 consist of FS and IFS spectroscopy \citep{Boker2022, Jakobsen2022}, obtained using the medium- (FS) and high-resolution (IFS) grating/filter combination G395/F290LP, delivering a spectral resolving power of $R \sim 1000$ and $R \sim 2700$, and continuous wavelength coverage from 2.87 to 5.27\,$\mu$m. J0217-0208 was observed with FS mode with a total on-source exposure time of 3107\,s using the S200A2 slit ($0.2^{\prime \prime} \times 3.3^{\prime \prime}$ size). Data were retrieved from MAST and calibrated using the CRDS context version \texttt{11.17.19} and file version \texttt{1256}.
The flux density of the FS spectrum was matched to that obtained from photometry in NIRCam F356W to account for slit-losses. 
IFS observations, totalling 1750\,s exposure time, were calibrated with the JWST pipeline release \texttt{1.18.0} \citep{Bushouse2025zndo}, using the CRDS context version \texttt{12.1.8} and file version \texttt{1364}. In addition to the standard three-stage reduction, custom steps are introduced to optimize data quality (e.g., \citealt{Marshall2023, Ubler2023, Perna2023}). First, we enable the \texttt{msa\_flagging} and \texttt{outlier\_detection} modules, which are skipped by default by the pipeline, while the \texttt{ipc} correction step is intentionally omitted. Then, we remove the 1/f noise, which produces vertical striping on the detector images \citep{Perna2023}. Finally, we apply a sigma-clipping routine to the count-rate images to generate a temporal bad/hot pixel mask, flagging pixels as \texttt{`DO\_NOT\_USE'} if identified as outliers in 3 out of 4 dithers. The final spectral cube was produced with the \texttt{drizzle} algorithm. Figure \ref{fig3} shows the FS spectra and the collapsed IFS H$\beta$+[O\,{\sc iii}]\,$\lambda\lambda$4960,5008 image of J0217-0208.

\subsection{Emission line measurements}\label{section_gauss}
\looseness=-1

Given the superior SNR of the NIRSpec/FS spectrum compared to the IFS one, we use it to measure the emission line fluxes of J0217-0208. The spectrum reveals strong emission in H$\beta$, [O{\sc iii}]\,$\lambda\lambda$4960,5008, and H$\alpha$, as well as in several other lines that are typically faint, such as [S~{\sc ii}]\,$\lambda\lambda$6718,6732 and the temperature-sensitive [O{\sc iii}]\,$\lambda$4363 line (detected with a $\simeq 6\sigma$ significance). The continuum emission is also well detected across the spectral coverage of the G395M grating, with a SNR per pixel ranging from $\sim$ 3 to 9.
Using the H$\beta$ and [O{\sc iii}]\,$\lambda\lambda$4960,5008 emission lines, we determine the systemic redshift of $z_{\rm sys} = 6.2038 \pm 0.0001$. 
\looseness=-1

\begin{table}
\begin{center}
\caption{Summary of the properties of J0217-0208. \label{table1}}
\begin{tabular}{l c c}
\hline \hline
\smallskip
\smallskip
Property  & Value & Uncertainty  \\
\hline 
R.A. (J2000)  & 02:17:21.59 & $0.1^{\prime \prime}$ \\
Dec. (J2000)  &  $-$02:08:52.6 & $0.1^{\prime \prime}$ \\
$z_{\rm sys}$   & 6.2038 & $0.0001$   \\
$M_{\rm UV}$ (AB)   & $-23.37$ & $0.08$ \\
$r_{\rm eff, \star}$ (pc)  & $266$ & $10$ \\
$r_{\rm eff,neb}$ (pc)  & $550$ & 340 \\
Age (Myr)  & $5.5$   &   $1.5$  \\
$\tau$ (Myr)  & $12.6$   &   $6.7$  \\
12+log(O/H) & 8.20 & 0.11 \\
log(N/O) & $-$0.30 & 0.10 \\ 
E(B-V)$_{\rm stellar,sys}$  & $0.01$   &   $0.01$   \\
E(B-V)$_{\rm neb,sys}$  & 0.06   &   0.05   \\
E(B-V)$_{\rm neb,outflow}$  & 0.57   &   0.26   \\
SFR ($M_{\odot}$~yr$^{-1}$) & $142$ & $24$   \\
log($M_{\star}/M_{\odot}$)   & $9.11$ & $0.07$ \\
sSFR (Gyr$^{-1}$)  & 110   & 15  \\ 
$\log(\Sigma_{M}$ / $M_{\odot}\, \rm pc^{-2}$) & 3.43 & 0.37  \\
$\log(\Sigma_{SFR}$ / $M_{\odot}\,\rm yr^{-1}\,kpc^{-2}$) & 2.50 & 0.08  \\

\hline 
\end{tabular}
\end{center}
\end{table}
\looseness=-1

We measure the fluxes, equivalent widths, and line widths of the emission lines by fitting each line with a single Gaussian component. As a first step, we model and subtract the underlying continuum using spectral windows of $\Delta \lambda_{\rm rest} = 50$\,\AA{} on either side of each emission line.
The fitting is performed using the \texttt{Python} package \texttt{curvefit}, where the amplitude and line width are treated as free parameters. The centroid of each Gaussian is initially set based on the expected observed wavelength, i.e., $\lambda_{\rm obs} = \lambda_{\rm rest} \times (1+z)$, with redshift $z$ also included as a free parameter. To allow for potential uncertainties in line centroids, we permit small deviations (up to 10\%) in the fitted central wavelengths.
Additionally, the relative flux ratios of the [N~{\sc ii}] and [O~{\sc iii}] doublets are fixed to 2.94 and 2.98, respectively. 
To estimate uncertainties, we repeat the fitting process on 500 simulated spectra, introducing random noise to the observed spectrum. The noise is drawn from a Gaussian distribution with a standard deviation set by the 1$\sigma$ uncertainty in the observed spectrum.
\looseness=-1

\begin{figure*}
  \centering
  \includegraphics[width=0.8\textwidth]{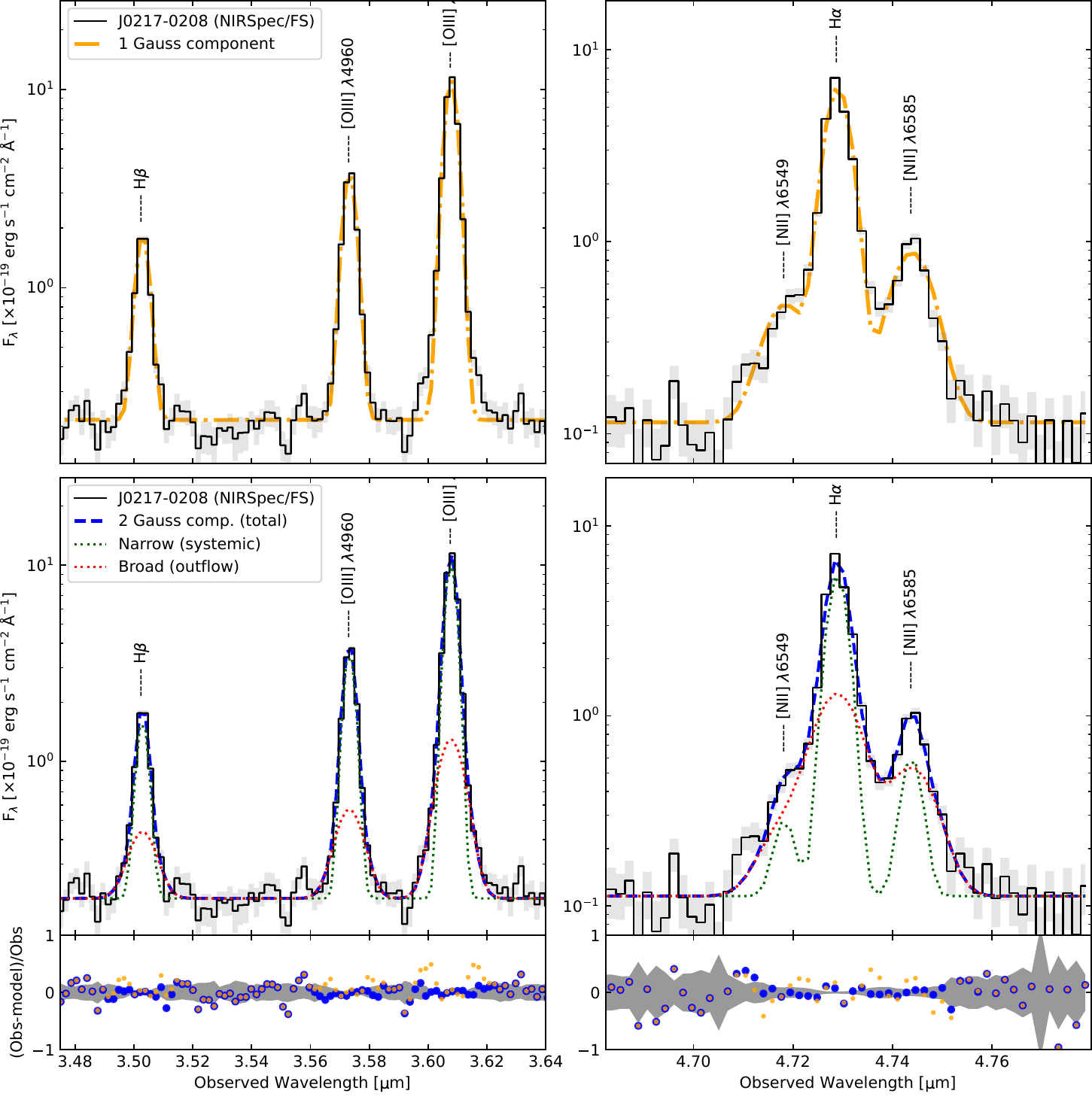}
  \caption{NIRSpec/FS spectrum of J0217-0208 (black) around H$\beta$ + [O~{\sc iii}] $\lambda\lambda$4960,5008 (left) and H$\alpha$ + [N~{\sc ii}] $\lambda\lambda$6549,6585 (right) emission lines. Top panels show the best-fit model assuming one Gaussian component for each emission line (orange). The bottom panel shows the best-fit model assuming two Gaussian components. The total emission is shown as a blue dashed line, while the individual components are shown as dotted lines (green and red for the narrow and broad components). Note that the vertical axes are in logarithmic scales. The gray area represents the $\pm 1 \sigma$ uncertainty spectrum. The lower panels show the residuals of the fits, with yellow and blue points corresponding to the single- and two-component models, respectively. }
  \label{fig4}
\end{figure*}

We find that a single Gaussian component generally provides a decent fit to the spectral profiles of most emission lines. The simultaneous fit of H$\beta$, [O{\sc iii}], H$\alpha$, and [N{\sc ii}] emission lines assuming a single Gaussian component yields a reduced $\chi^{2}_{\nu} = 13.0$ (with 6 free parameters).
The lines appear only marginally resolved, with observed line widths approximately 10-20\% larger than the expected instrumental widths. Thus, we assume conservatively that the lines are not fully resolved, and we do not correct them for the instrumental broadening. However, the brightest emission lines, H$\beta$, [O{\sc iii}]\,$\lambda\lambda$4960,5008, H$\alpha$, and [N{\sc ii}]\,$\lambda\lambda$6549,6585, exhibit prominent broad wings in their profiles, which are not well reproduced by the best-fit single Gaussian models (shown in orange in Figure \ref{fig4}). 
\looseness=-1

\looseness=-1

To better capture the core and wing components of the H$\beta$, [O{\sc iii}], H$\alpha$, and [N~{\sc ii}] lines, we repeat the fitting process using a two-component Gaussian model. The widths of the Gaussian components are constrained to remain consistent across different emission lines, although we take in consideration the variation of the NIRSpec G395M spectral resolution as a function of wavelength \citep{Rigby2023}. As done before, the relative flux ratios of the [N~{\sc ii}] and [O~{\sc iii}] doublets are fixed to 2.94 and 2.98, respectively, for both the narrow and broad components. 
\looseness=-1

The two-component fit (shown in blue in Fig.\,\ref{fig4}) provides a significantly improved match to the observed line profiles, particularly in the wings of the lines, with a reduced $\chi^{2}_{\nu} = 3.73$ (with 12 free parameters). We also calculate the Akaike Information Criterion (AIC) between the two-component and single-component fits, and find that the two-component fit is strongly preferred ($\Delta \rm AIC = 10.6$). The narrow component appears barely resolved only ($\Delta \lambda_{\rm narrow} = 19.4 \pm 3.5$\AA). 
On the other hand, the broad component exhibits a width of $\rm FWHM_{\rm broad} = 595 \pm 124$\,km\,s$^{-1}$. We also tested whether a broad component is required to fit the profiles of the fainter emission lines. In these cases, the broad component is not significantly detected, suggesting that the bulk of the emission in the faint lines is well described by a single (narrow) Gaussian component. The flux measurements for the total, narrow, and broad components are listed in Table \ref{table2}. 
\looseness=-1

From this point onward, we assume that the narrow component originates from warm gas tracing the systemic velocity due to virial motions within the galaxy. On the other hand, the broad component ($\rm FWHM_{\rm broad} = 595 \pm 124$ km s$^{-1}$, $\Delta v = -34 \pm 85$\,km\,s$^{-1}$) is identified with outflowing ionized gas, as it is detected in both Balmer (H$\beta$ and H$\alpha$) and metal lines ([O{\sc iii}], [N{\sc ii}]) and is required to explain their broad wings (Figure \ref{fig4}).
\looseness=-1

Finally, we also check the presence of a broad emission in H$\alpha$ that could be associated with a type-I AGN. We repeat the fitting process but adding a broad H$\alpha$ component with FWHM$>1000$\,km\,s$^{-1}$, hereafter the AGN component, in addition to the narrow and outflow components obtained in the previous fit. 
We fixed the central wavelengths and line widths of the narrow ($\Delta v = 0$ km s$^{-1}$, $\rm FWHM_{\rm narrow} = 290$ km s$^{-1}$) and outflow ($\Delta v = -34$ km s$^{-1}$, $\rm FWHM_{\rm broad} = 595$ km s$^{-1}$) components, while leaving all the amplitudes and the line width of the AGN component as free parameters. Our results indicate that the AGN component is not significantly detected, and we obtain an upper limit of $F_{\rm AGN} \rm (H\alpha) \leq 6.9 \times 10^{-18}$ erg s cm$^{-2}$ (3$\sigma$) assuming an AGN FWHM$=1000$\,km\,s$^{-1}$. Our results thus support the lack of type-I AGN activity in J0217-0208 (see also \citealt{Phillips2025}), or it is heavily obscured. 
\looseness=-1

\subsection{Differential dust attenuation and line diagnostics}\label{line_diagnostics}
\looseness=-1

Thanks to the high SNR of the FS/G395M spectrum of J0217-0208, we can constrain the nebular dust attenuation for the narrow and broad components independently. For the narrow component, we use the flux measurements of H$\gamma$ obtained from the single Gaussian fit, along with the narrow components of H$\beta$ and H$\alpha$ from the two-component fit. For the outflowing gas, we use the broad components of H$\beta$ and H$\alpha$.
\looseness=-1

We assume Case B recombination for $T_{\rm e} = 10^{4}$~K and $n_{\rm e} = 1000$ cm$^{-3}$ (see Section \ref{density}), and adopt the \cite{cardelli1989} extinction law with $R_{\rm V} = 3.1$. We find $E(B-V)_{\rm sys} = 0.06\pm0.05$ and $E(B-V)_{\rm out} = 0.57\pm0.26$ for the systemic and outflowing components, respectively, in excellent agreement with the values reported by \cite{Matsuoka2025}. These values are used to deredden the observed fluxes. For emission lines requiring a two-component fit (H$\beta$, [O{\sc iii}], H$\alpha$, and [N{\sc ii}]), their total dereddened fluxes are obtained by summing the dust-corrected fluxes of the narrow and broad components, each corrected using its respective $E(B-V)$. 
\looseness=-1

Using the total dust-corrected line fluxes, we investigate the nature of the ionizing source in J0217-0208. We first explore the classical Baldwin, Phillips \& Terlevich (BPT) diagnostics \citep{BPT, BPT2}, employing the line ratios [O{\sc iii}]/H$\beta$, [N{\sc ii}]/H$\alpha$, [S{\sc ii}]/H$\alpha$, and [O{\sc i}]/H$\alpha$. We obtain log([O{\sc iii}]/H$\beta$)$=0.71 \pm 0.08$, log([N{\sc ii}]/H$\alpha$)$=-0.59 \pm 0.07$, log([S{\sc ii}]/H$\alpha$)$=-1.59 \pm 0.05$, and log([O{\sc i}]/H$\alpha$)$\leq -2.14$ (3$\sigma$).
\looseness=-1

\begin{figure}
  \centering
  \includegraphics[width=0.39\textwidth]{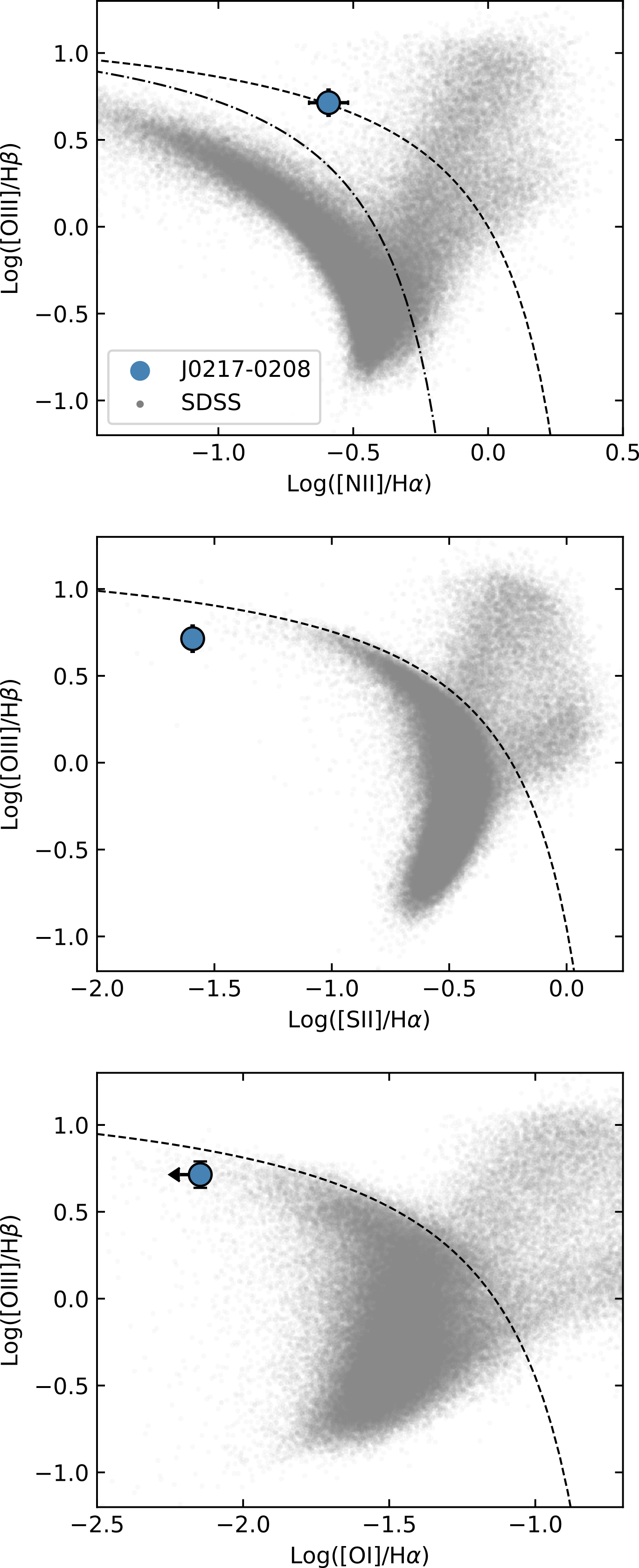}
    \caption{Location of J0217-0208 (blue) on the classic BPT diagnostic diagrams: [O{\sc iii}]/H$\beta$ vs. [N{\sc ii}]/H$\alpha$ (top), [S{\sc ii}]/H$\alpha$ (middle) and [O{\sc i}]/H$\alpha$ (bottom). The dot-dashed curve in the top panel shows the empirical star-forming/AGN demarcation from \cite{Kauffmann2003MNRAS.346.1055K}, while the dashed curves indicate the maximum starburst lines from \cite{Kewley2001} (top) and \cite{Kewley2006} (middle and bottom). Grey circles represent local SDSS galaxies for comparison.}
  \label{fig_BPT}
\end{figure}

As shown in Figure~\ref{fig_BPT} and according to \cite{Kewley2006}, the derived [O{\sc iii}]/H$\beta$ vs. [S{\sc ii}]/H$\alpha$ and [O{\sc i}]/H$\alpha$ line ratios are consistent with J0217-0208 being powered by star formation, placing it far from the locus of low-$z$ AGNs. However, its position in the [O{\sc iii}]/H$\beta$ vs. [N{\sc ii}]/H$\alpha$ diagram 
is more ambiguous, as it lies along the "maximum starburst" curve of \cite{Kewley2001}. This could suggest a hard stellar ionizing radiation field (cf. \citealt{steidel2014}), shocks contributing to the line excitation, or a possible non-negligible AGN contribution. Additionally, it may reflect an elevated nitrogen abundance in J0217-0208, similar to that observed in some UV-bright high-$z$ galaxies \citep{Marques2024, Senchyna2024, Schaerer2024, Topping2024}, which could enhance the [N{\sc ii}]/H$\alpha$ ratio \citep{Stiavelli2024, Zhang2025}.

From the non-detection of He{\sc ii} $\lambda$4686, we derive log(He{\sc ii}/H$\beta) < -1.61$. According to \cite{Shirazi2012}, this upper limit is consistent with J0217-0208 being powered by star formation, without requiring a hard ionizing source such as an AGN. Finally, we measure [O{\sc iii}]$\lambda$4363/H$\gamma = 0.20 \pm 0.03$, which aligns with star formation in all models presented by \cite{Mazzolari2024}.
\looseness=-1

In short, nearly all line diagnostics examined here indicate that J0217-0208 is a star-forming galaxy with negligible contribution from an AGN.
\looseness=-1

\subsection{Electron density, temperature, and abundances}\label{density}
\looseness=-1

For the ISM properties, we do not attempt to derive separate physical conditions for the systemic and outflow components, but instead focus on the luminosity-weighted properties of the ionized gas. This is because the temperature-sensitive [O {\sc iii}] $\lambda4363$ line and the density-sensitive [S {\sc ii}] $\lambda\lambda6716,6731$ doublet are only significantly detected in the narrow component, preventing independent measurements of $T_{\rm e}$ and $n_{\rm e}$ for the broad component.\footnote{Recently, \cite{harikane2025} found evidence for an inhomogeneous ISM structure in J0217-0208 from a relatively strong far-infrared [O~{\sc iii}] 88$\mu$m emission that cannot be explained by uniform electron density and temperature. However, their results should not significantly impact our measurements of densities, temperatures, and abundances.} We therefore adopt the electron temperature and density derived from the narrow component and apply them to the total dust-corrected line fluxes when deriving abundances.

\looseness=-1

Within the spectral coverage of G395M, the [S\,{\sc ii}] $\lambda\lambda$6716/6731 doublet is the only $n_{\rm e}$-sensitive spectral feature detected with high significance. 
We obtain [S\,{\sc ii}]\,$\lambda$6716/$\lambda$6731 $= 0.88 \pm 0.22$ and estimate log($n_{\rm e}/$cm$^{-3}) = 3.06^{+0.33}_{-0.30}$  using the \texttt{getTemDen} task from Pyneb \citep{Luridiana2015}. 
Assuming log($n_{\rm e}/$cm$^{-3}) = 3.06$ and using the observed [O{\sc iii}]\,$\lambda$4363/$\lambda$5008 line ratio of ($7.6 \pm 0.1) \times 10^{-3}$, we measure $T_{\rm e}$(O{\sc iii}$) = (1.05 \pm 0.05) \times 10^{4}$~K. 
\looseness=-1

\begin{figure}
  \centering
  \includegraphics[width=0.48\textwidth]{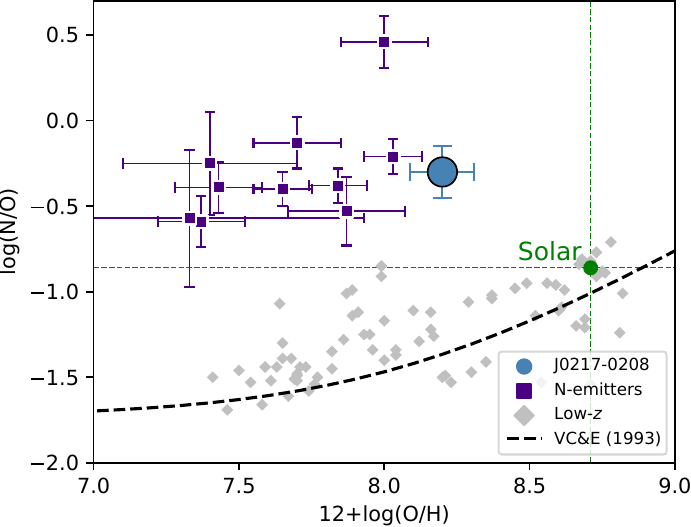}
  \caption{Chemical abundance ratio N/O as a function of O/H. J0217-0208 is shown as a blue circle, while other $z\gtrsim 5$ nitrogen emitters are marked by violet squares (compiled by \citealt{Schaerer2024}; see also \citealt{Ji2025} for a more extensive sample). For comparison, low-redshift star-forming galaxies and H\,{\sc ii} regions from \citet{Izotov2023} are plotted in grey. The dashed line shows the average trend in low-$z$ galaxies, as parameterized by \citet{VCE1993}. }
  \label{fig_NO}
\end{figure}

For abundances, we consider O$^{+}$ and O$^{2+}$ from [O~{\sc ii}] $\lambda\lambda$3727,3729 and [O~{\sc iii}] $\lambda$5008, respectively. While [O~{\sc ii}] is not covered by our G395M spectrum, we obtain its flux using the observed line ratio of [O~{\sc iii}]/[O~{\sc ii}] $=4.39\pm0.42$ from \cite{harikane2025}, and rescale and correct it for dust attenuation. Using the empirical relation of \cite{Izotov2006}, we derive an oxygen abundance $12 + \log(\rm O/H) = 8.20 \pm 0.11$, in excellent agreement with that obtained by \cite{harikane2025}. We also rely on \cite{Izotov2006} to derive the nitrogen over oxygen ratio by applying the ionization correction factor $\rm ICF (N^{+}) = 6.91$, obtaining $12+\log(\rm N/O) = -0.30$. As such, J0217+0208 has a sub-solar O/H ($\simeq 0.30\,Z_{\odot}$) and super-solar N/O ($\simeq 3.6 \times$ solar), as highlighted in Figure \ref{fig_NO}. 

It is also worth noting that the broad component of [N\,{\sc ii}] composes a larger fraction of its total flux ($\simeq 66\%$ or $\simeq 86\%$ after dust correction) than other lines (H$\beta$, [O\,{\sc iii}], H$\alpha$; see Fig.~\ref{fig4} and Section \ref{outflow}). While this may indicate enhanced nitrogen enrichment in the outflowing gas \citep[e.g.,][]{ArellanoCordova2025, Rizzuti2025}, the lack of independent density and temperature measurements for the broad component prevents us from deriving separate abundances.

\looseness=-1

\looseness=-1

\subsection{Outflow properties}\label{outflow}
\looseness=-1

From Section \ref{section_gauss}, we found that the spectral profiles of the H$\beta$, [O{\sc iii}], H$\alpha$, and [N{\sc ii}] emission lines require a narrow and broad components, the latter associated with ionized outflows. This component has a line width of $\rm FWHM_{\rm broad} = 595 \pm 124$ km s$^{-1}$ and is centered at $\Delta v = -34 \pm 85$ km s$^{-1}$ relative to the systemic velocity. The observed broad-to-narrow flux ratio varies between $f_{\rm out}/f_{\rm sys} \simeq 0.25$ for [O{\sc iii}] and $\simeq 0.49$ for H$\alpha$. After correcting for dust attenuation, these values increase by a factor of $\sim3-5$, reaching $f_{\rm out}/f_{\rm sys} \simeq 1.25-1.64$ due to different attenuation levels between the narrow and broad components. 
\looseness=-1

We derive an outflow velocity, defined as $v_{\rm out} = \vert \Delta v \vert + \rm FWHM_{\rm broad}/2$, of $v_{\rm out} = 331 \pm 135$\,km\,s$^{-1}$ (e.g., \citealt{Arribas2014}). 
To estimate the total mass of ionized gas in the outflow, $M_{\rm out}$, we use the H$\alpha$ luminosity as a tracer of the ionizing photon production rate ($Q_{\rm H}$), given that $M_{\rm out} \propto Q_{H}/n_{e}$, thus $M_{\rm out} \propto L(\mathrm{H}\alpha)/n_{e}$ (e.g., \citealt{Colina1991}). Using the dust-corrected luminosity of the outflowing component, $L(\mathrm{H}\alpha) = (3.02 \pm 0.4) \times 10^{43}$ erg s$^{-1}$, and assuming a constant ISM density of $n_{e} = 10^{3}$ cm$^{-3}$ (Section \ref{density}), we derive log($M_{\rm out}/M_{\odot}) = 8.07 \pm 0.07$. The ionized gas mass in the outflow constitutes approximately $62\%$ of the total (systemic + outflowing) ionized gas mass traced by H$\alpha$.
\looseness=-1

To determine the spatial extent of the outflow, $r_{\rm eff,out}$, we use NIRSpec/IFU observations of J0217-0208. Since the IFU data around H$\alpha$ is noisier and has lower spatial resolution than that around H$\beta$+[O{\sc iii}], we collapse the 2D emission of H$\beta$ and [O{\sc iii}] $\lambda\lambda$4960,5008 using narrow spectral windows of $\delta \lambda_{\rm rest} = \pm 6$\,\AA{ }centered on these lines. To remove the underlying continuum, we subtract a collapsed IFU spectrum on both sides of the H$\beta$+[O{\sc iii}] lines, covering $\lambda_{\rm rest} = 4400-4800$\,\AA{ }and $\lambda_{\rm rest} = 5050-5450$\AA. The resulting NIRSpec/IFU 2D image of H$\beta$+[O{\sc iii}] emission is shown in Figure \ref{fig3}.
\looseness=-1

We fit a 2D Gaussian profile to the H$\beta$+[O{\sc iii}] emission, while masking the region around the faint component $\approx 0.38^{\prime \prime}$ SW of J0217-0208. We find a circularized $\rm FWHM = 0.25^{\prime \prime} \pm 0.11^{\prime \prime}$. 
To correct for the instrumental PSF, we use the empirical NIRSpec PSF measurements as a function of wavelength provided by \cite{DEugenio2024}. At $\lambda = 3.5-3.6\mu$m, the NIRSpec/IFU PSF is around $\simeq 0.11^{\prime \prime}-0.14^{\prime \prime}$ (depending on each method, see: \citealt{DEugenio2024}) and we assume the intermediate value $\rm FWHM (PSF) = 0.125^{\prime \prime}$. Thus, the PSF corrected size of the H$\beta$+[O{\sc iii}] emission is $1.1 \pm 0.7$\,kpc (FWHM), or $r_{\rm eff,out} = 0.55 \pm 0.34$\,kpc assuming a Gaussian profile. This is consistent with the size derived using the NIRCam/F356W image of J0217-0208, which predominantly traces the H$\beta$+[O{\sc iii}] emission (Section \ref{morphology_photometry}). We thus assume a $r_{\rm eff,out} = 0.55 \pm 0.34$\,kpc for the extension of the outflow. We emphasize, however, that this value should be regarded as an assumption rather than a robust measurement, given the uncertainties in the NIRSpec PSF and the contribution of the narrow component to the ionized gas emission.
Considering this size, we obtain an outflowing gas rate, $\dot M_{\rm out} = M_{\rm out} \times v_{\rm out} \times r_{\rm eff,out}^{-1} = 72\pm49\,M_{\odot}$\,yr$^{-1}$, defined as the amount of gas expelled due to the outflow per unit time.
\looseness=-1

\subsection{Morphology and photometry}\label{morphology_photometry}
\looseness=-1

J0217-0208 shows a compact morphology in the NIRCam images, tracing the rest-UV to optical (top panels of Figure \ref{fig3}). Additionally, it has a much fainter companion located $\approx 0.38^{\prime \prime}$ southwest of J0217-0208 (or $\approx 2.1$\,kpc proper). A close inspection of the IFS data confirms the same redshift as J0217-0208, suggesting that this faint companion is likely a minor merger. However, we exclude the faint component in this morphological analysis.
\looseness=-1

We first investigate the light distribution of J0217-0208 using \texttt{PySersic} \citep{Pasha2023}. \texttt{PySersic} fits the morphology of a source using 2D S\'ersic models convolved with a given point-spread function (PSF) and uses a Bayesian framework to understand the degeneracies between different parameters. We generate PSFs for all eight NIRCam filters (F115W to F444W) using \texttt{WebbPSF}. 
We assume 2D S\'ersic profiles with a S\'ersic index varying from 0.5 to 6.0, while other parameters such as total flux, ellipticity, and orientation are left free. The fit is performed on a $60 \times 60$ px cutout centered on J0217-0208, masking the emission from the fainter companion (using a $0.2^{\prime \prime}$-radius aperture).
\looseness=-1

For the F150W image ($\lambda_{\rm eff,rest} \simeq 2000$\,\AA), we obtain an effective radius $r_{\rm eff,F150W} = 266 \pm 10$\,pc and a S\'ersic index $n=1.12 \pm 0.10$. On the other hand, the morphology in F356W, which traces predominantly the combined emission from H$\beta$+[O{\sc iii}] lines, shows a more extended profile, with $r_{\rm eff,F356W} = 575 \pm 29$\,pc and a S\'ersic index $n=1.78 \pm 0.11$, which is in excellent agreement with the size of the H$\beta$+[O{\sc iii}] emission derived from the NIRSpec/IFU observations (Section \ref{outflow}). 
\looseness=-1

For the remaining NIRCam bands, the measured sizes are consistent with previous estimates. J0217-0208 exhibits a relatively resolved morphology in bands tracing the stellar continuum, with sizes comparable to those obtained in F150W, albeit with larger uncertainties. In the F444W image, which probes H$\alpha$+[N{\sc ii}] emission, J0217-0208 shows a spatial extent similar to that in F356W, with $r_{\rm eff,F444W} = 646 \pm 107$\,pc. In summary, J0217-0208 shows a compact stellar emission with $r_{\rm eff,\star} = 266 \pm 10$\,pc, while its nebular emission appears more extended, with $r_{\rm eff,neb} = 575 - 650$\,pc.
\looseness=-1

Finally, we perform aperture photometry on all NIRCam images. We use relatively large apertures, varying from 0.45$^{\prime\prime}$ at shorter wavelengths (F115W) to 0.58$^{\prime\prime}$ at longer wavelengths (F444W), to account for the total emission of J0217$-$0208 and PSF variations across the NIRCam filters. Errors are derived using a circular annulus around the source to estimate the local background noise, but include the propagated uncertainties within the photometric aperture. Overall, J02017$-$0208 shows a steep rest-UV SED, as shown in Figure \ref{fig7}, with $f_{\nu} \rm (F115W) = 1.42 \pm 0.05 \mu$Jy, $f_{\nu} \rm (F150W) = 1.24 \pm 0.03 \mu$Jy, and $f_{\nu} \rm (F200W) =1.02 \pm 0.06 \mu$Jy, covering the spectral range $\lambda_{\rm rest} \simeq 1600-2800$\,\AA. A linear fit to these measurements yields a UV slope of $\beta_{\rm UV} = -2.58 \pm 0.07$. At longer wavelengths, J0217-0208 presents a relatively flat SED, except in NIRCam bands containing bright optical emission lines, H$\beta$+[O{\sc iii}] ($f_{\nu} \rm (F356W) = 1.78 \pm 0.06 \mu$Jy) and H$\alpha$+[N{\sc ii}] ($f_{\nu} \rm (F444W) = 1.44 \pm 0.05 \mu$Jy).
The aperture photometry agrees well with the fluxes obtained from the \texttt{PySersic} best-fit models, with differences all within $\leq 1.5\sigma$ for all filters (and with a mean of $\simeq 0.9\sigma$).
\looseness=-1

\subsection{Global properties from SED analysis}\label{SED}
\looseness=-1

We perform SED fitting using the CIGALE code \citep[V.2022.1;][]{Burgarella2005, Boquien2019}, incorporating photometry from F115W to F444W.
The star formation history (SFH) is modeled with two components: one describing the young stellar populations responsible for the observed steep UV slope ($\beta_{\rm UV} = -2.58 \pm 0.07$), and another accounting for a potential underlying mature stellar population that could contribute to the emission at longer wavelengths. The young stellar component follows an exponentially declining SFH, with $\tau$ and age varying from 1 to 20 Myr in 1 Myr steps. The older component assumes a single-burst model with ages from 100 to 500 Myr in 100 Myr steps.
\looseness=-1

We adopt stellar population models from \cite{bruzual2003}, assuming a \cite{chabrier2003} IMF and metallicities of $Z = 0.2 - 0.4$\,$Z_{\odot}$, based on our analysis in Section~\ref{density}. The ionization parameter ranges from log($U$) = $-4$ to $-1$ in 1 dex steps.
Given the expected differential attenuation between the stellar continuum ($\beta_{\rm UV} \simeq -2.6$) and nebular emission, particularly from the outflowing component ($E(B-V)_{\rm out} \simeq 0.6$), we allow variations in the ratio of the color excess of the stellar continuum to that of the nebular gas, with $E(B-V){\star} / E(B-V)_{\rm nebular}$ ranging from 0.1 to 1.0 in 0.1 steps. We adopt the Milky Way dust extinction law from \cite{cardelli1989} with $R_{\rm V} = 3.1$ as the dust attenuation law for the nebular emission and the \cite{calzetti2000} attenuation law for the stellar emission.
\looseness=-1

\begin{figure}
  \centering
  \includegraphics[width=0.45\textwidth]{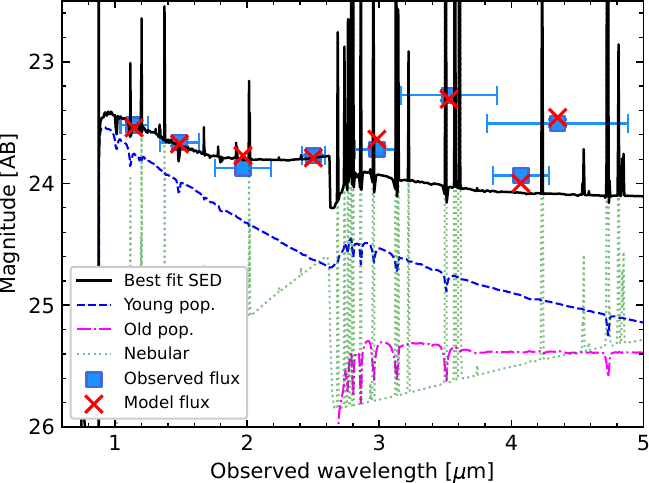}
  \caption{Best-fit SED model (black line) of J0217-0208 using NIRCam photometry (blue squares). The SED is dominated by the emission of a young stellar population characterized by an exponentially declining SFH with an age of $5.5\pm1.5$\,Myr ($\tau = 12.6\pm6.7$\,Myr), $\rm SFR = 142\pm24\,M_{\odot}$\,yr$^{-1}$, and a stellar mass $\log (M_{\star}^{\rm young}/M_{\odot})=9.11\pm0.07$. The nebular emission and the underlying old stellar population are represented by green and pink lines, respectively. }
  \label{fig7}
\end{figure}
\looseness=-1

Figure \ref{fig7} shows the best-fit model (black). The emission of J0217-0208 is dominated by a young stellar population in the whole spectral range covered by the NIRCam imaging data (dashed blue line), and has an age of $5.5\pm1.5$\,Myr and $\tau = 12.6\pm6.7$\,Myr. It has a 10~Myr-weighted $\rm SFR=142\pm24$\,$M_{\odot}$~yr$^{-1}$ and a stellar mass $\log (M_{\star}^{\rm young}/M_{\odot})=9.11\pm0.07$. We find a stellar color excess of $E(B-V)_{\star}=0.05\pm0.02$ with a $E(B-V)_{\star} / E(B-V)_{\rm nebular} = 0.6\pm0.3$. On the other hand, the old stellar population is not significantly detected and its mass should be less than $\log (M_{\star}^{\rm old}/M_{\odot}) \leq 9.5$ ($3\sigma$, magenta in Figure \ref{fig7}). 
From these results, we obtain a $\rm sSFR = 110 \pm 15$\,Gyr$^{-1}$ assuming only the starburst mass, or  $\rm sSFR \geq 32 $\,Gyr$^{-1}$ (3$\sigma$) when the stellar mass of the young and old stellar components are considered. Finally, using the size of $r_{\rm eff,\star}$ derived from the F150W image, we obtain a stellar mass and SFR surface densities of $\log (\Sigma_{M_{\star}} / M_{\odot}\,\rm pc^{-2}) = 3.43\pm0.37$ and $\log (\Sigma_{SFR} / M_{\odot}\,\rm yr^{-1}\, kpc^{-2}) = 2.50\pm0.08$, respectively, which are $\sim 100\times$ higher than that of typical galaxies at similar redshift \citep[e.g.,][]{Calabro2024}.
\looseness=-1

\section{Discussion}\label{section4}
\looseness=-1

\subsection{UV-bright galaxies as young, powerful starbursts with enhanced production of ionizing photons}\label{disc_1}
\looseness=-1

The young age inferred from the N\,{\sc v} P-Cygni profile in the stacked SHELLQs-LAE spectrum, $6.3^{+9}_{-3}$\,Myr under the assumption of constant star formation (Section \ref{nature-shellqs}), implies that the bulk of star formation in these systems has taken place within the past $\lesssim 15$\,Myr. This effectively rules out a dominant contribution to the UV luminosity from more extended ($\gg 15$\,Myr) star formation episodes, indicating that SHELLQs-LAEs are undergoing recent and powerful bursts of star formation. The high specific SFR of one of the SHELLQs-LAEs studied here, J0217-0208 with $\rm sSFR = 110 \pm 15$\,Gyr$^{-1}$ (or $\rm sSFR \geq 32 $\,Gyr$^{-1}$ if the total mass is considered), corroborates this picture. Furthermore, JWST observations of the few other SHELLQs-LAEs reveal blue UV slopes, $\beta_{\rm UV} \simeq -3.0$ \citep{Matsuoka2025}, indicative of very young ($< 5$\,Myr) star-bursting phases. 
\looseness=-1

\begin{figure}
  \centering
  \includegraphics[width=0.48\textwidth]{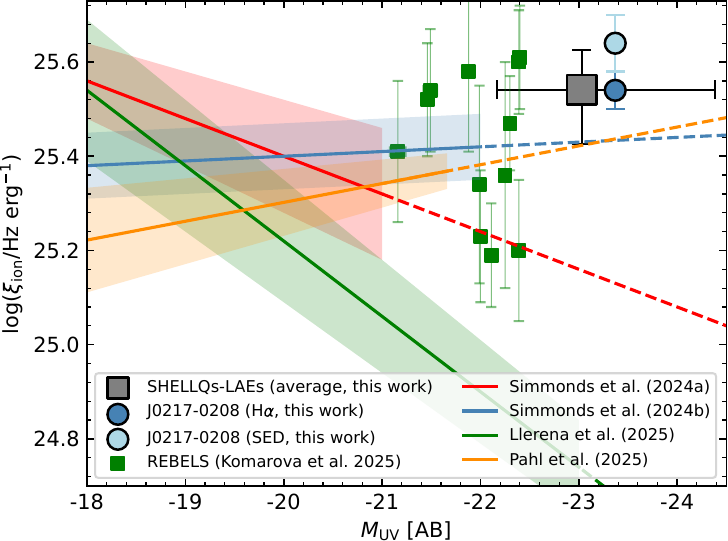}
  \caption{The ionizing photon production efficiency ($\xi_{\rm ion} = Q_{\rm ion} / L_{\nu}^{\rm UV}$) as a function of $M_{\rm UV}$. SHELLQs-LAEs (square; horizontal error-bar represents the $M_{\rm UV}$ coverage) show an elevated $\xi_{\rm ion}$, compared to the measurements (solid lines) and extrapolations (dashed lines) from \cite{Simmonds2024a, Simmonds2024b, Llerena2025}, and \cite{Pahl2025}. The $\xi_{\rm ion}$ of J0217-0208 derived from the H$\alpha$ luminosity and from the best-fit SED (all assuming $f_{\rm esc}=0$) are shown as a dark and light blue circles, respectively. Green squares represent the measurements of the $z\simeq 6-7$ REBELS UV-bright galaxies \citep{Komarova2025arXiv251110743K}. }
  \label{fig8}
\end{figure}

These findings imply that UV-bright SHELLQs-LAEs are prolific producers of ionizing photons. We estimate their average ionizing photon production efficiency, defined as $\xi_{\rm ion} = Q_{\rm ion} / L_{\nu}^{\rm UV}$, by integrating the ionizing emission ($Q_{\rm ion}$; i.e., photons with energies $>13.6$\,eV) from the best-fit BPASS synthetic spectrum and comparing it to the corresponding monochromatic UV luminosity ($L_{\nu}^{\rm UV}$, evaluated at 1600\,\AA{ }and assumed to be dust-free). We find $\log(\xi_{\rm ion}/\mathrm{Hz\,erg^{-1}}) = 25.54^{+0.09}_{-0.12}$ for the averaged population of SHELLQs-LAEs. For J0217-0208, we use its H$\alpha$ dust-corrected luminosity, $L_{\rm H\alpha} = 3.67\times 10^{43}$\,erg\,s$^{-1}$, as a proxy for $Q_{\rm ion}$, and measure $\log(\xi_{\rm ion,H\alpha}/\mathrm{Hz\,erg^{-1}}) = 25.53\pm0.04$ assuming zero escape fraction of ionizing photons (Figure~\ref{fig8}, but see also Section~\ref{disc_2}). For comparison, the best-fit SED of J0217-0208 predicts a $\log(\xi_{\rm ion,SED}/\mathrm{Hz\,erg^{-1}}) = 25.64 \pm 0.06$, i.e., only slightly higher ($\simeq 0.11$\,dex) than that derived using H$\alpha$ luminosity.
\looseness=-1

Figure~\ref{fig8} shows the average $\xi_{\rm ion}$ obtained for the SHELLQs-LAEs (square), which is approximately 0.34\,dex higher than the pre-JWST canonical value reported by \citet{robertson2015}.
For comparison, we also include in Figure~\ref{fig8} measurements of $\xi_{\rm ion}$ and its relationship with $M_{\rm UV}$, derived for fainter star-forming galaxies at $z \sim 3-9$ from recent \textit{JWST} studies (\citealt{Simmonds2024a, Simmonds2024b, Llerena2025}, and \citealt{Pahl2025}; shown in red, blue, green, and orange, respectively, in Figure~\ref{fig8}). While these studies have established $\xi_{\rm ion} - M_{\rm UV}$ relations for UV-fainter galaxies (solid lines in Figure~\ref{fig8}), we extrapolate those trends to the higher UV luminosities characteristic of the SHELLQs-LAEs (dashed lines).
Overall, these extrapolated relations do not fully account for the elevated $\xi_{\rm ion}$ observed in UV-bright systems like the SHELLQs-LAEs. That said, the extrapolations from \citet{Simmonds2024b} and \citet{Pahl2025} yield values only slightly lower, around $\log(\xi_{\rm ion}/\mathrm{Hz\,erg^{-1}}) \simeq 25.4$, but are consistent within our 1$\sigma$ uncertainties. In contrast, the extrapolated relation from \citet{Llerena2025} predicts significantly lower values, with $\log(\xi_{\rm ion}/\mathrm{Hz,erg^{-1}}) \lesssim 24.7$ at $M_{\rm UV} \lesssim -23$, corresponding to a factor of $\sim7$ lower than what we infer for the SHELLQs-LAEs. For comparison, Figure~\ref{fig8} also shows the $\xi_{\rm ion}$ measurements of UV-bright galaxies at $z\sim 6-7$ from the REBELS survey (green squares; \citealt{Komarova2025arXiv251110743K}).
\looseness=-1

\looseness=-1

However, potential biases should be considered and discussed. SHELLQs-LAEs are selected for relatively strong Ly$\alpha$ emission ($L_{\rm Ly\alpha} > 10^{43}$\,erg\,s$^{-1}$; see Section~\ref{selection}), which may bias the sample toward younger stellar populations with higher $\xi_{\rm ion}$. SHELLQs has also identified similarly UV-bright galaxies with weaker Ly$\alpha$ emission (shown in light blue in Figure~\ref{fig1}), whose stacked spectra have been analyzed in previous works \citep[e.g.,][]{Harikane2020, matsuoka2022}. However, these spectra show strong damped Ly$\alpha$ absorption that extends into the N~{\sc v}\,$\lambda$1240 region, preventing reliable estimates of stellar age and $\xi_{\rm ion}$. Whether these Ly$\alpha$-faint counterparts exhibit similarly high $\xi_{\rm ion}$ remains uncertain and requires further observations.
Another source of uncertainty is the stellar metallicity, which remains unconstrained. However, its impact on $\xi_{\rm ion}$ is expected to be minor: at fixed stellar age, reducing metallicity from 0.5\,$Z_{\odot}$ to 0.05\,$Z_{\odot}$ increases $\xi_{\rm ion}$ by only $\sim$0.1\,dex \citep{chisholm2019}.
Finally, our best-fit BPASS models assume a standard IMF. Recent evidence \citep{Upadhyaya2024} suggests that UV-bright galaxies at intermediate redshifts may host very massive stars (VMS; $M_{\star} > 100\,M_{\odot}$), via strong and broad He~{\sc ii}~$\lambda$1640 emission in their spectra \citep{crowther2016, Martins2022, Martins2025}. This could suggest an extended IMF upper mass cutoff $M_{\rm up}^{\rm IMF} > 100\,M_{\odot}$ in SHELLQs-LAEs, and their $\xi_{\rm ion}$ could be boosted by factors of $\sim 1.5-2.0$ \citep{Schaerer2025}.
\looseness=-1

\subsection{Strong dusty outflows in J0217-0208}\label{disc_2}
\looseness=-1

There is compelling evidence for strong, dusty outflows in J0217-0208. First, our analysis of the H$\beta$, [O~{\sc iii}]\,$\lambda\lambda$4960,5008, H$\alpha$, and [N{\sc ii}]\,$\lambda\lambda$6549,6585 emission line profiles (Section \ref{section_gauss}) clearly reveals a broad component with $\rm FWHM = 595 \pm 124$\,km\,s$^{-1}$  indicative of high-velocity outflowing gas. 
Despite known degeneracies in multi-component line fitting, which are accounted for in the uncertainties, the broad component shows a high level of obscuration, with $E(B-V)_{\rm out} = 0.57 \pm 0.26$, significantly higher than that inferred for the systemic narrow-line component, $E(B-V)_{\rm sys} = 0.06 \pm 0.05$ (see also: \citealt{DelPino2024}, \citealt{Crespo2025arXiv251114658C}, \citealt{Parlanti2025}). Furthermore, the stellar continuum of J0217–0208 appears nearly dust-free. Using the UV slope $\beta_{\rm UV} = -2.58 \pm 0.07$ measured from NIRCam photometry and adopting the \citet{calzetti2000} extinction law with an intrinsic (dust-free) slope of $\beta_{\rm UV,0} = -2.616$ \citep{reddy2018}, we estimate a stellar reddening of $E(B-V)_\star = 0.01 \pm 0.01$.
\looseness=-1

\begin{figure}
  \centering
  \includegraphics[width=0.45\textwidth]{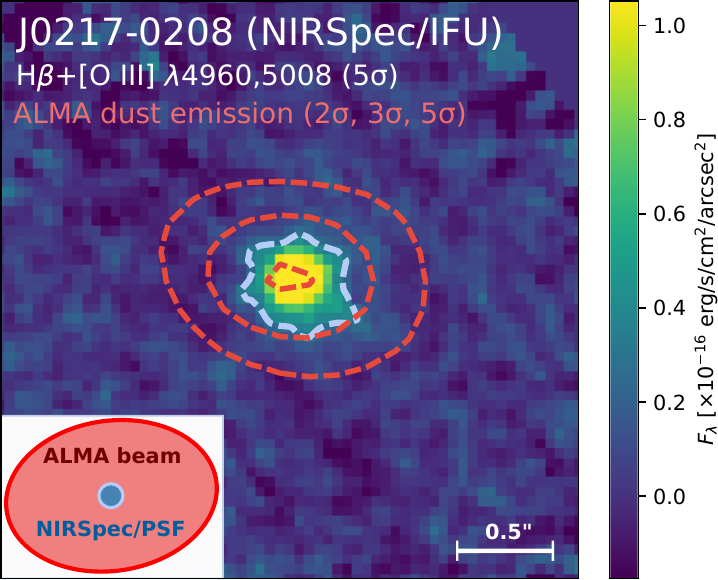}
  \caption{Collapsed H$\beta$ and [O\,{\sc iii}] $\lambda\lambda$4960,5008 map of J0217-0208 from NIRSpec/IFU (5$\sigma$ contour shown in dashed blue). Dust continuum emission from ALMA Band 7 is overlaid in red contours (2$\sigma$, 3$\sigma$, 5$\sigma$). The ALMA beam and the NIRSpec/IFU PSF are shown in the lower-left corner. }
  \label{fig6}
\end{figure}
\looseness=-1

Further support for dusty outflows comes from ALMA observations presented by \citet{Harikane2020b}, who detected dust continuum emission at 120\,$\mu$m and 160\,$\mu$m. From these measurements, they inferred a dust mass $M_{\rm dust} \simeq 1.8 \times 10^{8}\,M_\odot$ and a dust temperature $T_{\rm dust} \simeq 25$\,K (but see \citealt{Algera2024}). If all dust were confined within the stellar component of J0217–0208 ($r_{\rm eff,\star} \simeq 266$\,pc), the optical depth at 1500\,\AA\ would be extreme, of the order of $\tau_{1500} \approx 5000$ following \cite{Ziparo2023}, and no detectable UV and optical emission from J0217-0208 would be expected. This contradicts observations. Therefore, dust must be physically decoupled from the UV-bright stellar component. 
\looseness=-1

Altogether, our results strongly suggest the presence of strong outflows, carrying and pushing out dust well beyond the stellar emission. Our re-analysis of the ALMA data (Band 7, Figure~\ref{fig6}) shows that the dust emission in J0217–0208 is spatially resolved, with an observed major and minor axis of ($1.29^{\prime \prime}\pm 0.12^{\prime \prime}) \times (0.87^{\prime \prime}\pm 0.06^{\prime \prime}$) and a beam size $1.10^{\prime\prime} \times 0.78^{\prime\prime}$ (FWHM). We derive an effective dust radius of $r_{\rm eff, dust} = 1.37 \pm 0.44$\,kpc, which is substantially more extended than the starlight distribution. 
This could also explain the relatively low dust temperature measured in J0217–0208 ($T_{\rm dust} \simeq 25$\,K, \citealt{Harikane2020b}), which is considerably lower than that for galaxies at similar redshifts ($T_{\rm dust} \gtrsim 40$\,K; e.g., \citealt{Sommovigo2022, Mitsuhashi2024}), and could result from dust being pushed to the outskirts where the radiation field is weaker. In the following, we investigate possible mechanisms and implications for such dusty outflows. 
\looseness=-1

Radiative outflows have been recently proposed to explain the overabundance of UV-bright galaxies at $z>10$ and their observed steep UV slopes \citep{Ferrara2023, Fiore2023, Ferrara2024, Nakazato2024, Ferrara2025}. In this so-called "Attenuation-Free Model" (AFM), strong outflows are driven by radiation pressure onto dust and gas through super-Eddington phases expected in the early phases of a starburst. 
Specifically, \citet{Fiore2023} and \citet{Ferrara2023} suggest that radiation-driven outflows are efficiently triggered once the specific star formation rate exceeds a critical threshold of $\mathrm{sSFR} > 25$\,Gyr$^{-1}$ (or $\mathrm{sSFR} > 12$\,Gyr$^{-1}$ following \citealt{Nakazato2024}). This condition is clearly satisfied in J0217–0208 ($\mathrm{sSFR} \geq 32$\,Gyr$^{-1}$ at 3$\sigma$; see Section~\ref{SED} for details). However, given the rather extreme compactness and high surface densities of J0217–0208, additional conditions may be required to effectively drive a radiative outflow \citep[e.g.,][]{Menon2023}. Following \cite{Ziparo2023}, and given the $\log (\Sigma_{SFR} / M_{\odot}\,\rm yr^{-1}\, kpc^{-2}) \simeq 2.50$ measured for J0217–0208, radiation pressure exceeds gravitational pressure when the burstiness parameter $k_{\rm s} \geq 12$, which quantifies the deviation from the canonical Kennicutt-Schmidt relation (such that $\Sigma_{SFR} \propto k_{\rm s} \Sigma_{\rm gas}^{1.4}$). This translates to an upper limit on the gas surface density of $\log (\Sigma_{\rm gas}/ M_{\odot}\, \rm pc^{-2}) \leq 3.58$. Assuming gas and stars have the same size, this implies a high star-formation efficiency of $\epsilon_{\rm SF} \geq 0.43$ (assuming $\epsilon_{\rm SF} = M_{\star}/[M_{\star}+M_{\rm gas}]$) for an efficient radiation-driven outflow. This is somewhat consistent with the numerical radiation hydrodynamic simulations of \cite{Menon2023}, who find super-Eddington phases in similarly compact systems once high $\epsilon_{\rm SF}$ are reached (see also: \citealt{Menon2024b}). The extreme starbursting nature of J0217–0208, able to form $M_{\star} \approx 10^{9}\,M_{\odot}$ in just $\approx 6$\,Myr, and its elevated surface densities further suggests that enhanced/high star formation efficiencies are likely at play \citep{Dekel2023}. 
Therefore, high star formation efficiencies and radiation-driven outflows are likely causally linked, as suggested by recent works \citep{Dessauges-Zavadsky2025, Somerville2025}. 
\looseness=-1

An alternative explanation is mechanical feedback from supernova (SNe) explosions. 
Unlike radiative-driven outflows, SNe-driven feedback is inherently delayed by a certain amount of time after the burst of star formation ($\simeq 3 - 10$\,Myr), depending on the upper mass limit for SNe, which is still unconstrained \citep[e.g.,][]{Smartt2015}. If all O-type stars explode as SNe, we expect significant mechanical feedback right after the explosion of the most massive stars ($\sim 100\,M_{\odot}$), i.e., $\simeq 3$\,Myr after the onset of star formation. Considering the age derived for J0217–0208 ($5.5\pm1.5$\,Myr), this would allow a window of $\sim 1-4$\,Myr for significant SNe feedback. On the other hand, if only lower-mass O-type stars ($<20\,M_{\odot}$) explode as SNe \citep{Sukhbold2016}, the onset of mechanical feedback could be delayed by $\gtrsim 7$\,Myr, making significant SNe-driven outflows unlikely at the current evolutionary stage of J0217–0208. However, these estimates are simplified and do not account for cumulative SNe from previous ($\gg 10$\,Myr) episodes of star formation, which are likely required to explain the large dust mass observed in J0217–0208.
\looseness=-1

Whether mechanically or radiatively driven, the strong and heavily obscured outflow observed in J0217–0208 appears essential for clearing dust along its line of sight. Because gas and dust are the main sources of LyC opacity, these outflows may also facilitate the escape of a significant fraction of ionizing photons in J0217–0208, given its steep UV slope ($\beta_{\rm UV} = -2.58$; \citealt{chisholm2022}), but relatively weak nebular emission ($EW_{0}\,\rm (H\beta) \simeq 57$\,\AA; \citealt{zackrisson2013}), the small Ly$\alpha$ offset relative to the systemic velocity ($\Delta v \simeq 250$\,km\,s$^{-1}$; \citealt{izotov2018b}), and the high SFR surface density ($\log (\Sigma_{SFR} / M_{\odot}\,\rm yr^{-1}\, kpc^{-2}) \simeq 2.50$; \citealt{naidu2020}), all pointing to a LyC escape fraction between $f_{\rm esc}^{\rm LyC} \approx 18\%$ and $\approx 75\%$ depending on each indicator. 
\looseness=-1

\subsection{Comparison with literature and final remarks on UV-bright systems}
\looseness=-1

UV-bright systems like SHELLQs-LAEs ($M_{\rm UV} < -22.0$) are rare at any redshift. Here, we compare the properties of SHELLQs-LAEs with other UV-bright sources discovered in wide-area surveys. However, a direct quantitative comparison is challenging because the available datasets differ in several aspects, including wavelength coverage, depth, and spectral resolution, which can bias the inferred physical properties.
\looseness=-1

The BoRG survey \citep[e.g.,][]{Trenti2011} targeted luminous ($M_{\rm UV} \sim -21$) galaxies at $z \sim 8$ identified from HST pure-parallel observations. Follow-up JWST observations of a subset of these sources (marked in red in Figure~\ref{fig1}) have revealed steep UV slopes, with all (20) but three showing $\beta_{\rm UV} < -2.0$, consistent with negligible dust extinction \citep{roberts2024_borg}. Some BoRG sources, including the brightest ones ($M_{\rm UV} \lesssim -22.0$), exhibit Ly$\alpha$ emission in their NIRSpec/PRISM spectra \citep{roberts2024_borg}, although no quantitative measurements on Ly$\alpha$ properties are available so far. Furthermore, the star-formation histories derived by \citet{Rojas-Ruiz2025} suggest that BoRG galaxies are undergoing major bursts of star formation, consistent with the results obtained for SHELLQs-LAEs. 
\looseness=-1

Another well-studied population of UV-bright galaxies is provided by the REBELS survey \citep[][with $-21.5 < M_{\rm UV} < -23.0$; green in Figure \ref{fig1}]{bouwens2022}, which identified relatively massive systems ($\log (M_{\star}/M_{\odot}) \simeq 8.5-10.0$) with short light-weighted ages and high sSFRs, particularly among the UV-brightest members ($\mathrm{sSFR} \simeq 30$\,Gyr$^{-1}$ for $M_{\rm UV} < -22.0$ systems; \citealt{Topping2022_REBELS}). While REBELS galaxies also show relatively steep UV slopes ($\beta_{\rm UV} \sim -2.0$; \citealt{bouwens2022}, \citealt{Bowler2024_REBELS}, \citealt{Fisher2025MNRAS.539..109F}), they contain, on average, substantial dust reservoirs ($\log (M_{\rm dust}/M_{\odot}) \simeq 7.0-7.5$; \citealt{Dayal2022_REBELS}), suggesting that UV- and far-IR–emitting regions of, at least, some of these sources are not co-spatial \citep{Ferrara2022_REBELS}, a scenario that is also seen in the SHELLQs-LAE J0217–0208 (see Section~\ref{disc_2}). Whether this segregation is a consequence of strong feedback, as indicated for J0217–0208, remains unclear, but it could partly explain the relatively strong and redshifted Ly$\alpha$ emission detected in several REBELS UV-bright sources \citep{Endsley2022_REBELS}. 
\looseness=-1

At lower redshifts, UV-bright systems with comparable or even higher luminosities have been discovered in the wide-area SDSS/BOSS survey \citep[][]{marques2020b, marques2021, marques2022, Marques-Chaves2024b, Upadhyaya2024, Dessauges-Zavadsky2025}. These systems at $z \sim 2-4$ are even brighter ($-23.0 < M_{\rm UV} < -24.5$) than their higher-$z$ counterparts, and are characterized by powerful starbursts with young ages ($\lesssim 10$\,Myr) and high specific SFRs ($\sim 100$\,Gyr$^{-1}$), as revealed by strong stellar winds in rest-frame UV spectra (Figure \ref{fig2}-B2). Similar to REBELS sources and J0217–0208, these $z \sim 2-4$ UV-bright galaxies also contain large dust masses, $\log (M_{\rm dust}/M_{\odot}) \simeq 7.0$–$8.0$ \citep{Dessauges-Zavadsky2025}. In many cases, dust emission appears more extended (by factors of $\sim 2-5$) than the UV-optical distribution, which itself shows negligible dust obscuration ($\beta_{\rm UV}$ ranging from $-1.84$ to $-3.50$). Evidence for strong outflows is, however, scarce, being detected in only a few cases \citep{Alvarez2021, marques2021}. Moreover, ALMA observations of their molecular gas reveal short depletion timescales ($t_{\rm depl} \simeq 10-70$\,Myr) and high star-formation efficiencies ($\epsilon_{\rm SF} = 8\%$ to $>40\%$), indicating that these starbursts are caught at the very beginning of stellar mass build-up, during a phase of efficient gas-to-stars conversion \citep{Dessauges-Zavadsky2025}. 
\looseness=-1

\begin{figure}
  \centering
  \includegraphics[width=0.48\textwidth]{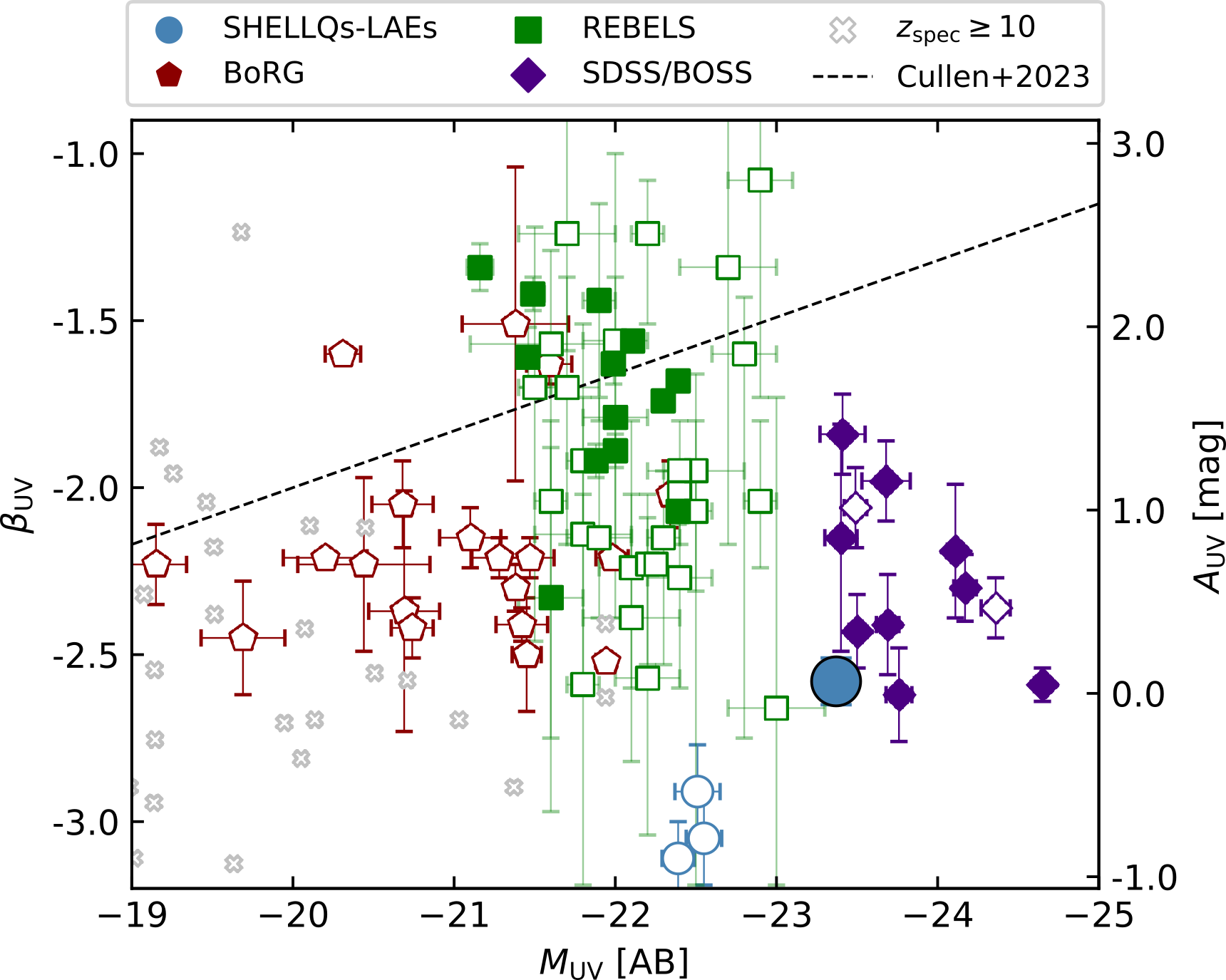}
  \caption{UV slope ($\beta_{\rm UV}$) as a function of $M_{\rm UV}$ for SHELLQs-LAEs at $z\sim 6$ (circles; this work and \citealt{Matsuoka2025}), BORG galaxies at $z\sim 8$ (pentagons; \citealt{roberts2024_borg}), REBELS sources at $z \sim 6.5-8.5$ (squares; \citealt{bouwens2022, Fisher2025MNRAS.539..109F}), UV-bright systems from SDSS/BOSS at $z\sim 2-4$ (diamonds; \citealt{Dessauges-Zavadsky2025}), and $z\geq 10$ sources confirmed with JWST. Sources with detected dust emission are shown with solid symbols, while sources without dust detection or with no dust observations are shown with empty symbols. Dashed line represents the $\beta_{\rm UV} - M_{\rm UV}$ relationship of \cite{Cullen2023} derived from fainter sources.}
  \label{fig8_final}
\end{figure}
\looseness=-1

Figure \ref{fig8_final} compares the UV slopes and $M_{\rm UV}$ of several SHELLQs-LAEs (circles) with those of BoRG (pentagons), REBELS (squares), and lower-redshift UV-bright galaxies selected from SDSS/BOSS (diamonds). Some of these sources have detections of dust emission (solid symbols), with dust-to-stellar mass ratios ranging from $\log \xi_{\rm d} \simeq -2.8$ to $\simeq -0.9$ \citep{Harikane2020b, Dayal2022_REBELS, Dessauges-Zavadsky2025}. 
Despite this, many exhibit steep UV slopes, consistent with little ($\beta_{\rm UV} \sim -2.0$, or $A_{\rm UV} \sim 1.0$\,mag) to negligible ($\beta_{\rm UV} \lesssim -2.5$, or $A_{\rm UV} \sim 0$\,mag) dust obscuration, as also seen for some of the UV-brightest systems at $z>10$ (grey in Figure \ref{fig8_final}). This behavior resembles the scenario proposed by \cite{Ferrara2023}, in which strong outflows clear dust along the line of sight, a feature that is observed in J0217-0208 through its powerful and heavily obscured ionized outflow.
\looseness=-1

Taken together, these populations of UV-bright galaxies suggest a common physical picture in which UV-luminous systems trace short-lived but intense starburst phases, characterized by rapid stellar mass assembly, efficient ionizing photon production, and complex dust–gas geometries. In light of recent JWST discoveries of unexpectedly UV-luminous galaxies at $z>10$, our results emphasize the interplay between very young stellar populations (e.g., \citealt{Roberts2025}), powerful outflows, and likely elevated star formation efficiencies, all of which act to boost the UV luminosities. As a final remark, some of the extreme properties inferred for J0217-0208, such as its steep UV slope, high SFR and stellar mass surface densities, and nitrogen enhancement, closely resemble those of the most UV-luminous systems at $z>10$, suggesting a shared mode of star formation and chemical enrichment. 
Future detailed studies of larger samples of UV-bright galaxies, including SHELLQs-LAEs and sources to be identified in upcoming wide-field surveys such as \textit{Euclid} and \textit{Roman} \citep[e.g.,][]{Weaver2025_EUCLID}, will be crucial to firmly establish these connections.
\looseness=-1

\section{Conclusions}\label{section5}
\looseness=-1

In this work, we have investigated the properties of 27 very UV-bright ($-22.0 \lesssim M_{\rm UV} \lesssim -24.4$), Ly$\alpha$ emitters ($L_{\rm Ly\alpha} > 10^{43.0}$\,erg\,s$^{-1}$) at $z \sim 6$, drawn from the Subaru High-$z$ Exploration of Low-Luminosity Quasars (SHELLQs) survey. In the first part of this study, we focused on the average rest-frame UV properties of these sources, referred to as SHELLQs-LAEs. By stacking individual optical spectra from ground-based observations, we constructed a composite rest-UV spectrum and obtained the following key results:
\looseness=-1

\begin{itemize}
    \item The stacked UV spectrum of SHELLQs-LAEs (Figure~\ref{fig1}) exhibits a narrow Ly$\alpha$ emission line ($\simeq 600$\,km\,s$^{-1}$ FWHM) and a prominent N~{\sc v} $\lambda1240$ P-Cygni profile, characterized by blueshifted absorption and redshifted emission. Comparison with BPASS stellar population synthesis models suggests that this feature is well reproduced by a very young stellar population with an age of $6.3^{+9}_{-3}$\,Myr under the assumption of constant star formation. We investigated whether AGN activity could explain the observed spectral signatures, but conclude that their contribution to the UV luminosity or frequency in the sample is likely small. \vspace{0.5mm}

    \item These findings imply that the strong UV luminosity of these sources arises from the emission of a large number of massive and hot stars, making SHELLQs-LAEs highly efficient ionizing sources. We derive an average ionizing photon production efficiency of $\log(\xi_{\rm ion}/\mathrm{Hz\,erg^{-1}}) = 25.54^{+0.09}_{-0.12}$, significantly higher than canonical pre-JWST values and extrapolations from UV-fainter galaxies observed with JWST. UV-bright systems like SHELLQs-LAEs are therefore intense starbursts, capable of forming at least $\approx 10^{9}\,M_{\odot}$ of stars within $<15$\,Myr, modulo dust attenuation.

\end{itemize}

In the second part of this work, we analyzed JWST observations of a representative SHELLQs-LAE, J0217–0208, at $z = 6.204$ with $M_{\rm UV} = -23.4$. The dataset comprises deep imaging in eight NIRCam filters (from F115W to F444W), as well as NIRSpec integral field and fixed-slit spectroscopy with the G395M grating ($\lambda = 2.8$–$5.2\,\mu$m, $R \sim 1000$). From these observations, we derived the following results:

\begin{itemize}

    \item Based on NIRCam photometry, we obtain a best-fit SED characterized by a young starburst with an age of $5.5 \pm 1.5$\,Myr (with an e-folding timescale $\tau = 12.6 \pm 6.7$\,Myr), a star formation rate of $\mathrm{SFR} = 142 \pm 24\,M_{\odot}$\,yr$^{-1}$, and a stellar mass of $\log (M_{\star}^{\rm young}/M_{\odot}) = 9.11 \pm 0.07$. The total stellar mass of J0217–0208 can be as high as $\log (M_{\star}^{\rm total}/M_{\odot}) \leq 9.65$ (3$\sigma$), owing to the presence of a significant old stellar population that is outshined by the powerful starburst. J0217-0208 thus presents a high specific star-formation rate of $\rm sSFR = 110 \pm 15$\,Gyr$^{-1}$, or  $\rm sSFR \geq 32 $\,Gyr$^{-1}$ (3$\sigma$) when the total stellar mass is considered. \vspace{0.5mm}

    \item J0217–0208 exhibits a compact morphology in the NIRCam bands tracing the rest-frame UV continuum, with an effective radius of $r_{\rm eff} = 266 \pm 10$\,pc. However, it appears moderately more extended in the F356W and F444W bands, where nebular emission from H$\beta$+[O~{\sc iii}] and H$\alpha$+[N~{\sc ii}] contributes significantly, respectively. We measure high stellar mass and SFR surface densities of $\log (\Sigma_{M_{\star}} / M_{\odot}\,\mathrm{pc}^{-2}) = 3.43 \pm 0.37$ and $\log (\Sigma_{SFR} / M_{\odot}\,\mathrm{yr}^{-1}\,\mathrm{kpc}^{-2}) = 2.50 \pm 0.08$, respectively.\vspace{0.5mm}

    \item The NIRSpec spectrum reveals a high S/N detection of both the stellar continuum and several optical emission lines. 
    Several line diagnostics suggest that J0217–0208 is powered by star formation without any strong evidence of a dominant AGN. Using the [S{\sc ii}] $\lambda\lambda$6716/6731 doublet, we measured a relatively high electron density, log($n_{\rm e}/$cm$^{-3}) = 3.06^{+0.33}_{-0.30}$. From the $T_{e}$-sensitive [O{\sc iii}] $\lambda$4363 line, we measure $T_{\rm e}$(O{\sc iii}$) = (1.05 \pm 0.05) \times 10^{4}$~K, for which we obtained $\log(\rm O/H) = 8.20\pm 0.11$. J0217–0208 also shows a supersolar nitrogen abundance of $\log(\rm N/O) = -0.30$. \vspace{0.5mm}
    
    \item The brightest rest-optical lines (H$\beta$, [O~{\sc iii}], H$\alpha$, [N~{\sc ii}]) require a broad kinematic component with $\mathrm{FWHM} = 595 \pm 124$\,km\,s$^{-1}$, indicative of high-velocity outflowing gas. This broad component is heavily obscured, with $E(B-V)_{\rm out} = 0.57 \pm 0.26$, in contrast to the nearly dust-free systemic narrow component, $E(B-V)_{\rm sys} = 0.06 \pm 0.05$, and stellar continuum, $E(B-V)_{\star} = 0.01 \pm 0.01$), which shows a steep UV slope of $\beta_{\rm UV} = -2.58 \pm 0.07$.   \vspace{0.5mm}

    \item Our results suggest that powerful dusty outflows are key to simultaneously explain several (a priori inconsistent) properties of J0217–0208, including its nearly dust-free stellar emission and, at the same time, the large dust mass ($\log (M_{\rm dust}/M_{\odot}) \simeq 8.3$), spatial extension ($r_{\rm eff,dust}\simeq 1.4$\,kpc), and low temperature ($T_{\rm dust} \simeq 25$\,K) inferred from ALMA observations. Whether mechanically or radiatively driven, these outflows appear capable of expelling dust and gas well beyond the stellar core, enabling J0217–0208 to appear exceptionally UV-bright. Such strong feedback may also have important implications, including a non-negligible escape of ionizing photons ($f_{\rm esc}^{\rm LyC} \approx 18-75\%$). 

\end{itemize}

In short, our results indicate that UV-luminous SHELLQs-LAEs, and likely other similar UV-bright sources, represent intense starburst phases characterized by efficient ionizing photon production and rapid stellar mass growth within only a few Myr. The case of J0217–0208 illustrates how young stellar populations with high specific star formation rates, feedback-driven outflows, and potentially enhanced star-formation efficiencies can act together to amplify the observed UV luminosities. Moreover, some of the most extreme properties derived for J0217–0208, including supersolar N/O at low O/H, steep UV slope, compact morphologies, and very high SFR and stellar mass surface densities, closely resemble those of the UV-brightest galaxies at $z>10$, suggesting a shared chemical enrichment and galaxy formation pathway. Systematic studies of larger samples of UV-bright galaxies, such as SHELLQs-LAEs and future discoveries from next-generation wide-field surveys with \textit{Euclid} and \textit{Roman}, will be essential to establish whether these mechanisms are ubiquitous among such extreme systems and to evaluate their broader role in early galaxy formation, evolution, and cosmic reionization.

\begin{acknowledgements}
We thank the anonymous referee for useful comments. 
This work is based on observations made with the NASA/ESA/CSA James Webb Space Telescope. The data were obtained from the Mikulski Archive for Space Telescopes at the Space Telescope Science Institute, which is operated by the Association of Universities for Research in Astronomy, Inc., under NASA contract NAS 5-03127 for JWST. These observations are associated with programs \#1657, 1840, and 1967.
Some of the data products presented herein were retrieved from the Dawn JWST Archive (DJA). DJA is an initiative of the Cosmic Dawn Center (DAWN), which is funded by the Danish National Research Foundation under grant DNRF140. 
J.A.-M., L.C., C.B.-P., L.C., and S. A. acknowledge support by grant CSIC/BILATERALES2025/BIJSP25022.
J.A.M. and C.B.P. acknowledge support by grant PID2024-158856NA-I00 from the Spanish Ministry of Science and Innovation/State Agency of Research MCIN/AEI/10.13039/501100011033 and by “ERDF A way of making Europe”.
L.C., J.A.M., and S.A. acknowledge support by grant PIB2021-127718NB-100 from the Spanish Ministry of Science and Innovation/State Agency of Research MCIN/AEI/10.13039/501100011033 and by “ERDF A way of making Europe”. 
C.B.P. acknowledges the support of the Consejer\'ia de Educaci\'on,Ciencia y Universidades de la Comunidad de Madrid through grants No. PEJ-2021-AI/TIC-21517 and PIPF-2023/TEC29505.
N.Y. acknowledges financial support from JSPS Specially Promoted Research 24H00004.
M.H. is supported by Japan Society for the Promotion of Science (JSPS) KAKENHI Grant No. 22H04939.
Y.W.R. is supported by JSPS KAKENHI Grant Number 23KJ2052.
K.M. is supported by JSPS KAKENHI grant Number 20K14516. K.M. and A.K.I are supported by JSPS KAKENHI grant No. 23H00131.
Y.N. is supported by JSPS KAKENHI Grant Number 23KJ0728 and a JSR fellowship.
D.C. is supported by research grant PID2021-122603NB-C21 funded by the Ministerio de Ciencia, Innovaci\'{o}n y Universidades (MI-CIU/FEDER) and the research grant CNS2024-154550 funded by MI-CIU/AEI/10.13039/501100011033.
A.C.G. acknowledges support by JWST contract B0215/JWST-GO-02926.
T.H. was supported by Leading Initiative for Excellent Young Researchers, MEXT, Japan (HJH02007) and by JSPS KAKENHI grant Nos. 22H01258, 23K22529, and 25K00020. M.O. is supported by the Japan Society for the Promotion of Science (JSPS) KAKENHI Grant Number 24K22894.

\end{acknowledgements}

\bibliographystyle{aa}
\bibliography{adssample_v2.bib}

\onecolumn

\begin{appendix}

\section{SHELLQs-LAEs sample}\label{appendix1}

\begin{table*}[h!]
\begin{center}
\caption{Summary of SHELLQs-LAEs. \label{table_sample}}
\begin{tabular}{l c c c c c c c c }
\hline \hline
\smallskip
\smallskip
ID  & Ra & Dec & $z$ & $M_{\rm UV}$ & FWHM (Ly$\alpha$) & $EW_{0}$\,(Ly$\alpha$) & $L$\,(Ly$\alpha$) & Notes  \\
    & (J2000) & (J2000) &  & [AB] & [km\,s$^{-1}$] & [\AA] & [erg\,s$^{-1}$] &  \\
\hline 

J0204$+$0015  & 02:04:15.74 & $+$00:15:34.5 & 6.043  &  $-24.16\pm0.03$  &  $610\pm80$  &  $16\pm2$  &  $43.76\pm0.05$  &  1  \\
J0217$-$0208  & 02:17:21.59 & $-$02:08:52.6 & 6.204  &  $-23.37\pm0.03$  &  $<230$  &  $15\pm1$  &  $43.33\pm0.04$  &  2,3,4,5  \\
J0220$-$0432  & 02:20:29.71 & $-$04:32:03.9 & 5.898  &  $-22.17\pm0.10$  &  $<230$  &  $29\pm2$  &  $43.15\pm0.03$  &  6  \\
J0235$-$0532  & 02:35:42.42 & $-$05:32:41.6 & 6.089  &  $-23.01\pm0.05$  &  $270\pm30$  &  $41\pm2$  &  $43.68\pm0.02$  &  2  \\
J0853$+$0139 & 08:53:48.84 & $+$01:39:11.0 & 6.007  &  $-22.51\pm0.14$  &  $<230$  &  $79\pm6$  &  $43.80\pm0.01$  &  2,3  \\
J0857$+$0056  & 08:57:38.53 & $+$00:56:12.7 & 6.343  &  $-23.01\pm0.07$  &  $620\pm90$  &  $57\pm5$  &  $43.85\pm0.02$  &  2  \\
J0905$+$0300  & 09:05:44.65 & $+$03:00:58.9 & 6.273  &  $-22.55\pm0.11$  &  $250\pm40$  &  $82\pm6$  &  $43.89\pm0.02$  &  2,3  \\
J0938$+$0154  & 09:38:30.24 & $+$01:54:05.8 & 6.304  &  $-22.91\pm0.15$  &  $300\pm360$  &  $70\pm12$  &  $43.90\pm0.05$  &  1  \\
J0947$+$0349  & 09:47:13.88 & $+$03:49:39.8 & 6.241  &  $-22.77\pm0.14$  &  $750\pm350$  &  $150\pm20$  &  $44.19\pm0.01$  &  1  \\
J1044$+$0102  & 10:44:29.18 & $+$01:02:07.1 & 6.200  &  $-23.29\pm0.07$  &  $370\pm420$  &  $11\pm2$  &  $43.16\pm0.08$  &  1  \\
J1107$+$0411  & 11:07:46.24 & $+$04:11:01.2 &  5.984 &  $-23.18\pm0.10$  &  $400\pm60$  &  $15\pm2$  &  $43.20\pm0.04$  &  1  \\
J1132$+$0038  & 11:32:18.15 & $+$00:38:00.1 & 5.654  &  $-23.18\pm0.05$  &  $650\pm130$  &  $24\pm4$  &  $43.34\pm0.04$  &  7  \\
J1159$+$0200  & 11:59:24.01 & $+$02:00:41.8 &  6.130  &  $-22.31\pm0.12$  &  $810\pm460$  &  $57\pm7$  &  $43.58\pm0.02$  &  1  \\
J1202$+$0256  & 12:02:53.13 & $+$02:56:30.8 & 6.021  &  $-22.78\pm0.14$  &  $240\pm180$  &  $28\pm4$  &  $43.56\pm0.03$  &  1  \\
J1207$-$0005  &  12:07:54.14 & $-$00:05:53.3 & 6.010  &  $-22.77\pm0.08$  &  $420\pm120$  &  $9\pm2$  &  $42.92\pm0.03$  &  8  \\
J1209$-$0006  &  12:09:24.01 & $-$00:06:46.5 & 5.848  &  $-22.51\pm0.05$  &  $580\pm50$  &  $26\pm5$  &  $43.01\pm0.04$  &  6  \\
J1258$-$0047  & 12:58:45.66 & $-$00:47:57.5 &  6.220 &  $-22.99\pm0.11$  &  $220\pm150$  &  $34\pm7$  &  $43.70\pm0.03$  &  1  \\
J1350$-$0027  & 13:50:12.04 & $-$00:27:05.2 &  6.474  &  $-24.38\pm0.19$  &  $620\pm200$  &  $56\pm11$  &  $44.39\pm0.04$  &  7  \\
J1400$-$0011  & 14:00:28.80 & $-$00:11:51.4 & 6.029  &  $-22.95\pm0.11$  &  $620\pm60$  &  $56\pm3$  &  $43.84\pm0.01$  &  6  \\
J1416$+$0015  & 14:16:12.71 & $+$00:15:46.2 & 6.027  &  $-22.39\pm0.10$  &  $230\pm20$  &  $98\pm5$  &  $43.86\pm0.01$  &  2,3  \\
J1417$+$0117  & 14:17:28.67 & $+$01:17:12.4 &  6.020  &  $-22.83\pm0.05$  &  $420\pm70$  &  $11\pm1$  &  $43.06\pm0.03$  &  2  \\
J1440$-$0107  & 14:40:01.30 & $-$01:07:02.2 & 6.121  &  $-22.59\pm0.10$  &  $440\pm260$  &  $21\pm2$  &  $43.27\pm0.03$  &  2  \\
J1450$-$0144  & 14:50:05.39 & $-$01:44:38.9 & 6.611  &  $-23.73\pm0.09$  &  $900\pm300$  &  $12\pm1$  &  $43.48\pm0.04$  &  1  \\
J2201$+$0155  & 22:01:32.07 & $+$01:55:29.0 &  6.157  &  $-22.97\pm0.04$  &  $320\pm70$  &  $24\pm1$  &  $43.46\pm0.01$  &  2  \\
J2228$+$0128  & 22:28:27.83 & $+$01:28:09.5 & 6.011  &  $-22.65\pm0.12$  &  $280\pm70$  &  $26\pm4$  &  $43.34\pm0.03$  &  8  \\
J2232$+$0012  & 22:32:12.03 & $+$00:12:38.4 &  6.181  &  $-22.81\pm0.34$  &  $300\pm120$  &  $120\pm16$  &  $44.04\pm0.02$  &  8,3  \\
J2252$+$0402  & 22:52:09.17 & $+$04:02:43.8 & 6.672  &  $-24.02\pm0.09$  &  $<230$  &  $12\pm1$  &  $43.48\pm0.03$  &  1  \\
J2304$+$0045  & 23:04:22.97 & $+$00:45:05.4 & 6.342  &  $-24.28\pm0.03$  &  $710\pm40$  &  $15\pm1$  &  $43.79\pm0.03$  &  6  \\

\hline 
\end{tabular}
\end{center}
\textbf{References ---} 1: \cite{matsuoka2022}; 2: \cite{matsuoka2018a}; 3: \cite{Matsuoka2025}; 4: \cite{harikane2025}; 5: this work; 6: \cite{matsuoka2018b}; 7: \cite{matsuoka2019}; 8: \cite{matsuoka2016}. 
\end{table*}

\newpage

\section{Emission Line measurements}
\label{BPTs}

\begin{table}[h!]
\begin{center}
\caption{Emission line measurements of J0217-0208. \label{table2}}
\begin{tabular}{l c c}
\hline \hline
\smallskip
\smallskip
Line  & Observed Flux & $EW_{0}$  \\
 & [$\times 10^{-18}$\,erg\,s$^{-1}$\,cm$^{-2}]$ & [\AA]  \\
\hline 

H$\delta$ $\lambda 4102$ & $ 3.05 \pm 0.35 $ & $ 10.4 \pm 1.4 $ \\

H$\gamma$ $\lambda 4341$ & $ 5.35 \pm 0.23 $ & $ 20.8 \pm 1.1 $ \\

[O{\sc iii}] $\lambda 4363$ & $ 1.08 \pm 0.17 $ & $ 4.2 \pm 0.7 $ \\

He{\sc ii} $\lambda 4686$ & $ <0.69 $ (3$\sigma$) & $ <2.1$ (3$\sigma$) \\

H$\beta$ $\lambda 4862$ & $ 12.89 \pm 1.08 $ & $ 56.8 \pm 5.4 $ \\

[O{\sc iii}] $\lambda 4960$ & $ 27.74 \pm 2.53 $ & $ 124.2 \pm 12.2 $ \\

[O{\sc iii}] $\lambda 5008$ & $ 82.11 \pm 3.64 $ & $ 380.3 \pm 18.4 $ \\

He{\sc i} $\lambda 5877$ & $ 2.03 \pm 0.18 $ & $ 22.04\pm 2.1 $ \\

[O{\sc i}] $\lambda 6302$ & $ <0.66 $ (3$\sigma$) & $ <3.0 $ (3$\sigma$) \\

[N{\sc ii}] $\lambda 6550$ & $ 3.23 \pm 0.41 $ & $ 44.8 \pm 6.2 $ \\

H$\alpha$ $\lambda 6564$ & $ 52.82 \pm 3.63 $ & $ 473.7 \pm 33.7 $ \\

[N{\sc ii}] $\lambda 6585$ & $ 9.49 \pm 1.25 $ & $ 96.2 \pm 13.3 $ \\

[S{\sc ii}] $\lambda 6718$ & $ 1.12 \pm 0.19 $ & $ 9.7 \pm 1.8 $ \\

[S{\sc ii}] $\lambda 6732$ & $ 1.27 \pm 0.23 $ & $ 11.0 \pm 2.1 $ \\

\hline 

H$\beta$ $\lambda 4862$ (narrow) & $ 9.37 \pm 0.73 $ &  --- \\

H$\beta$ $\lambda 4862$ (broad) & $ 3.52 \pm 0.79 $ &  --- \\

[O{\sc iii}] $\lambda$5008 (narrow) & $65.94 \pm 2.73$ & --- \\

[O{\sc iii}] $\lambda$5008 (broad) & $16.17 \pm 2.42$ & --- \\

H$\alpha$ $\lambda 6564$ (narrow) & $ 35.21 \pm 2.54 $ & --- \\

H$\alpha$ $\lambda 6564$ (broad) & $ 17.62 \pm 2.59 $ &  ---\\

[N{\sc ii}] $\lambda$6585 (narrow) & $3.19 \pm 0.80$ & ---\\

[N{\sc ii}] $\lambda$6585 (outflow) & $6.32\pm0.97$ & ---\\

\hline 
\end{tabular}
\end{center}
\end{table}

\end{appendix}

\end{document}